\documentstyle[prl,aps,multicol,epsf]{revtex}
\renewcommand{\Re}{\mbox{Re}}
\renewcommand{\Im}{\mbox{Im}}
\newcommand{\bea}{\begin{eqnarray}}
\newcommand{\eea}{\end{eqnarray}}
\newcommand{\beq}{\begin{equation}}
\newcommand{\eeq}{\end{equation}}
\newcommand{\sign}{\mbox{\em sign}}
\newcommand{\la}{\langle}
\newcommand{\ra}{\rangle}

\newcommand{\up}{\uparrow}
\newcommand{\down}{\downarrow}

\newcommand{\St}{\tilde{S}}
\newcommand{\Unity}{\large{1}}
\renewcommand{\imath}{i}

\newcommand{\tphi}{\tilde{\phi}}

\begin{document}
\draft
\tighten
\title{Interplay of superconductivity and magnetism in strong coupling.}
\author{C. N. A. van Duin and J. Zaanen}
\address{Institute Lorentz for Theoretical Physics, Leiden University\\
P.O.B. 9506, 2300 RA Leiden, The Netherlands}
\date{\today ; E-mail:cvduin@lorentz.leidenuniv.nl; jan@lorentz.leidenuniv.nl}
\maketitle
\begin{abstract}
A model is introduced describing the interplay between superconductivity and spin-ordering. 
It is characterized by on-site repulsive electron-electron interactions, 
causing antiferromagnetism, and nearest-neighbor attractive interactions,
giving rise to d-wave superconductivity. Due to a special choice for the
lattice, this model has a strong-coupling limit where the 
superconductivity can be described by a bosonic theory, similar to
the strongly coupled negative $U$ Hubbard model. This limit is analyzed
in the present paper. A rich mean-field
phase diagram is found and the leading quantum corrections to the mean-field
results are calculated. The first-order line between the antiferromagnetic- and
the superconducting phase is found to terminate at a tricritical point, where two second-order
lines originate. At these lines, the system undergoes a transition to- and 
from a phase
exhibiting both antiferromagnetic order and superconductivity. At finite
temperatures above the spin-disordering line,
quantum-critical behavior
is found. For specific values of the model parameters, it is possible to obtain 
$SO(5)$ symmetry involving the spin- and the phase-sector at the tricritical
point. Although this symmetry is explicitly broken by the projection to the
lower Hubbard band, it survives on the mean-field
level, and modes related to a spontaneously broken $SO(5)$ symmetry are
present on  the level of the random phase approximation in the superconducting phase.
\end{abstract}
%71.27 +a Strongly correlated electron systems; heavy fermions
%74.72.-h HTC compounds
%75.10 -b ,Gen. theory and models of magnetic ordering
%74.25 DW (Superconductivity phase diagrams)

\begin{multicols}{2}

\section{Introduction}

Both for empirical- and historical reasons, research on superconductivity
tends to be preoccupied with the weak coupling limit. From a more general
perspective, BCS theory as well as Gorkov-Migdal-Eliashberg theory 
correspond with a special case which in a sense is pathological. The
emphasis is completely on the amplitude of the order parameter while
fundamentally superconductivity is about breaking of gauge symmetry,
associated with the phase sector. The work of Schmitt-Rink and
Nozieres \cite{NSM} revealed that the BCS theory for a s-wave superconductor can be
smoothly continued to the strong coupling limit. It is generally recognized
that it is far easier to understand the vacuum structure of such a
superconductor in strong coupling. Amplitude fluctuations can be regarded
as highly massive excitations and all what remains is the phase sector
described in terms of hard-core bosons, or alternatively in terms of
pseudo-spin models.

In the context of high $T_c$ superconductivity one encounters a far more
complex physics. Abundant evidence is available for a d-wave superconducting
order parameter. This is usually discussed in
terms of weak-coupling theory with its d-wave nodal fermions while the
more sophisticated approaches start from this limit, attempting to
penetrate the intermediate coupling regime using self-consistent 
perturbation theory \cite{dwave}. The obvious problem is that the coherence length
is rather short \cite{Fischer}. At the same time,
an interesting case has been presented claiming that much of the
thermodynamics can be understood from phase-dynamics alone \cite{phasefluc}, completely
disregarding amplitude fluctuations. It would therefore
be useful to study strong coupling theories for d-wave superconductors.

An even better reason to pursue a strong coupling perspective is the
growing evidence for the presence of well  developed antiferromagnetism
coexisting with the superconductivity.
Traditionally, this was approached within, again, an implicitly weakly
coupled perspective. The magnetic fluctuations as seen in NMR and
neutron scattering were believed to be due to the proximity to an
amplitude driven spin density wave transition \cite{SDW}. Recently, this perspective
has been drastically changed due to the observation
of strong static antiferromagnetic order associated with the stripe phases
in the $La_2CuO_4$ system \cite{Tranq}. In the $Nd$ doped samples where this
order is strongest the magnitude of the N\'eel order parameter can be
as large as 0.3 $\mu_B$ \cite{punt3}, while 0.1 $\mu_B$ has been claimed in `pristine'
$La_{1.88} Sr_{0.12}  Cu O_4$  \cite{Endoh}. It appears that this antiferromagnetic
order is in  competition or even coexisting with the superconducting order
\cite{Endoh,TranqPRL}.
Given that the stripe antiferromagnet should be strongly renormalized
downward due to transversal quantum spin fluctuations \cite{qSpins}
the stripe antiferromagnet has to be strongly coupled. Given the strong
similarities between the static order and the incommensurate spin
fluctuations which seem to be generic for all cuprate superconductors 
in the underdoped regime, a strong coupling perspective 
on the antiferromagnetism should be closer to the truth even if static order 
is not present, at least as long as the doping is not too large.

Recently several theoretical attempts have been undertaken 
to shed light on this
problem of strongly coupled superconductivity and antiferromagnetism. 
The simplest theory of this kind is Zhang's $SO(5)$ theory, where
superconductivity and antiferromagnetism are `unified' within a single
larger symmetry \cite{Zhangscience}. Given that no such symmetry is manifestly present
at the ultraviolet of the problem, this might well be misleading
and one would like to have a more general framework in which this
(near) $SO(5)$ symmetry appears as a special case. The manifest 
symmetry of the problem is $U(1) \times SU(2)$ (superconducting phase-
and spin, respectively). The structure 
of the long wavelength effective theory based on this symmetry
principle has been analyzed recently by one of the authors \cite{J}, 
including the charge order associated with the stripe phase. These
approaches are only truly meaningful at long wavelength and a
more complete understanding is in high demand. In fact, the only reasonably
complete theory is the one by Vojta and Sachdev \cite{VS}, based on the large 
$N$/small $S$ saddle point of the $Spl (2N)$ $t-J$ model. However, in
this large $N$ limit the antiferromagnetism is in the strongly quantum
disordered regime, and is therefore at best dual to the renormalized
classical N\'eel order of the stripe phases.

Here we will present an exceedingly simple toy model which seems 
nevertheless to catch much of the physics discussed in the above.
It is similar in spirit to the lattice-boson description of
superconductivity and magnetism discussed in  Ref's \cite{Eder} and \cite{pSO5}.
The pursuit is to construct a model which at the same time
describes localized magnetism and local pairing superconductivity.
The magnetism is undoubtedly related to strong, Hubbard U type
on-site repulsions. This prohibits for obvious reasons on-site paring.
The next microscopic length scale available on the lattice is the
lattice constant itself: the pairs causing the superconductivity live
on the links of the lattice \cite{realpairs}. If such a link-pair is occupied, the
sites connected by this link are both occupied by a single electron.
In the presence of on-site repulsions these electrons will tend to
turn into a spin system. The number fluctuations implied by the
superconducting phase order correspond with such an occupied link-pair
becoming unoccupied, causing at the same time a dilution of the spin
system.

On the square lattice a subtlety keeps a theory with these link pairs
as  building blocks from being simple. Different from the large $N$ 
limit with its spin-Peierls order \cite{VS}, the link pairs cause both conceptual
problems in describing the state at half-filling as well as serious
technical problems. As will be discussed in Section II,  a consistent formulation requires local constraints
to be added to the theory in order to exclude tilings of the lattice
characterized by multiple occupancies on the sites. This is not
necessarily fatal: the theory is bosonic and it might well be that
Jastrov projections cure the problem. A central result of this paper
is our discovery of a different lattice where these likely 
non-essential `correlation' problems are absent: the $1/5$ depleted
lattice shown in Fig. 1. The link-pairs live on the long bonds,
while the short bonds only carry spin-spin interactions. As will be further discussed, this model is characterized by an unproblematic classical
(in fact, large $d$) limit. This allows us to derive in a controlled way
a complete semi-classical description.

As discussed in Section III, we find a surprisingly rich phase diagram on
the classical level containing all phases, which have been up to now 
suggested in this context, including the large $N$ spin-quantum paramagnets.
Perturbing around this classical limit, we address the structure of
the semiclassical theory including the universality classes  at the
various phase transitions (Section IV). By fine tuning parameters, 
we find lines in the phase diagram where the $SO(5)$ symmetry is 
approached. However, even at the most symmetric point $SO(5)$ is not
reached: as we will show, the theory becomes $SO(5)$ symmetric on the
classical level but the quantum corrections destroy this symmetry again.
As was already pointed out in the context of the $SO(5)$ symmetric ladders,
fine tuning of the on-site repulsions is required to stabilize the full
symmetry (Sections V and VI).  

\section{The model}
\subsection{Correlated superconductivity}

For the strong-coupling description we are aiming at, the microscopic
building blocks are electron link-pairs, created by the operators
\beq
{L_{i,\delta}^{\sigma_1\sigma_2}}^{\dagger}=c_{i,\sigma_1}^{\dagger}c_{i+\delta,\sigma_2}^{\dagger} \, ,
\label{links}
\eeq
where $\delta$ is a lattice unit-vector, while $i$ labels the sites.
Such a link-pair is the typical microscopic object in a strong-coupling
theory of d-wave superconductivity and the smallest electron-pair that 
can support spin degrees of freedom. 
Two serious problems arise when trying to construct a model from these
operators, one technical and one conceptual.
The technical problem is related to the spatial structure of the link-pairs, 
which introduces correlations between pairs centered on different bonds. 
These correlations show up in the commutation relations of the link
operators. Operators along different bonds
do not commute if their links share a common site. As a result, the
dimension of the link-operator algebra grows with the system size.
This makes a simple pseudo-spin description of the charge-sector impossible
and not much seems to have been gained by
going to the strong-coupling limit.

This problem can be avoided by assuming that one can somehow keep track of
which electrons belong to a particular pair
(this can be ambiguous, for instance in the case of four electrons 
sitting in a square). 
If this is possible, the link-pairs can be described by hard-core boson-operators, satisfying
$b^{\sigma_1\sigma_2}_{i,\delta}
b^{\sigma_1\sigma_2\,\dagger}_{i,\delta'}=0$ for $\delta\neq\delta'$.
Link-bosons on different bonds always commute, removing the problem
of the infinite-dimensional link-algebra. The correlation effects then
show up in a different way, however.
The  hard-core link-bosons are spinful generalizations of the quantum dimers 
\cite{Kivelson} \cite{Fradkin}. It is well known that even the classical 
theory of the dimers is a complex combinatorics problem, which was solved
for the case of half-filling \cite{Kastelein}, but not for general densities.
This problem seems unavoidable when one tries to construct a strong-coupling
theory for electron-pairs with one or the other real space internal structure on the
square lattice.

The conceptual problem is related to the fact that our link-pairs carry spin.
It concerns the state at half-filling. 
On the square lattice, there are many ways in which the
link-pairs can be distributed over the lattice to obtain 
complete covering. Since the half-filling state is a pure spin-system, this
charge degree of freedom is superfluous. The link-pair model at half-filling
therefore suffers from a large degeneracy. 

In the
large-$N$ $t$-$J$ model studied by Vojta and Sachdev \cite{VS},
 link-pairing arises as a result of nearest-neighbor spin-singlet 
formation, and the pairs are in this case spin-zero dimers.
% In this case, the link-pairs can be identified by considering the spin-state
%and one could use a hard-core boson description of the pairs. 
As a result, different link-pair configurations at half-filling 
correspond to different distributions of the singlet spin-bonds over
the square lattice. These configurations are therefore physically distinct.
The spin-Peierls order which is present at half-filling singles out a 
particular link-pair configuration, breaking the degeneracy.

For a large $S$ type antiferromagnet, however, 
the spin-sector cannot be used to break the degeneracy associated with half-filling. 
Let us therefore consider a model where link-pairing arises as a result of charge-charge interactions.
In this case, link-pairs can have both a singlet and a triplet 
spin-component, allowing for the construction of a half-filling antiferromagnet.
Consider a nearest-neighbor attractive
interaction $V$, an on-site repulsive interaction $U$ and a 
longer-range repulsive interaction $U'$, 
\bea
{\cal H} & = & \sum_i\left[-V\sum_{\delta}n_i n_{i+\delta}+U n_{i\up}n_{i\down} \right.\nonumber \\
 & & \left.
+U'\sum_{\delta_1,\delta_2\neq-\delta_1}n_i n_{i+\delta_1+\delta_2} \right] + \mbox{hopping processes} \, ,
\eea
where $\delta$ runs over all lattice unit-vectors.
%The on-site interaction $U$ is needed to obtain antiferromagnetism. 
The attractive interaction $V$ promotes link-pairing, while
the longer range repulsive interaction $U'$ is needed to counteract phase-separation in the strong-coupling limit.

At small electron  densities, the strong-coupling limit of the above model 
describes a dilute gas of electron link-pairs. Near half-filling, it describes
a dilute gas of {\em hole} link-pairs, moving through a spin background.
Taking hole-pairs and spins, instead of electron-pairs, as the elementary 
building blocks in the strong-coupling limit near half-filling, the large degeneracy
in the description is avoided. Such a perspective is not entirely satisfactory,
however, since the spin-sector is in this case represented in a first-quantized
form. 

The technical problems, related to
the spatial correlations between the link-pairs, of course remain also for
this model. These correlations
become important at finite densities away from zero- or half-filling,
severely complicating the  strong-coupling analysis of this model.
Moreover, the short-range attractive and long-range repulsive interactions
will give rise to charge ordering phenomena at intermediate densities,
further complicating the physics.

\begin{figure}[t]
\epsfxsize=0.8\hsize
\hspace{.1 \hsize}
\epsffile{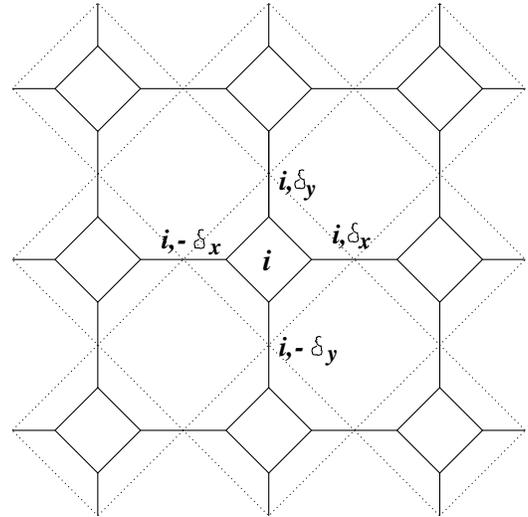} \vspace{8pt}
\caption{The $\frac{1}{5}$-depleted lattice. Dotted lines connect nearest-neighbor horizontal and vertical bonds.}
\label{fig het rooster}
\end{figure}

\subsection{Depleted lattice}

The complex spatial
correlations between link pairs and the tendency towards charge-ordering
 at intermediate densities as discussed in the previous subsection can be avoided  
by formulating 
the model not on the square lattice, but on the 1/5-depleted lattice, shown in 
Fig.\ref{fig het rooster}. We arrive at this lattice by expanding the sites
of a square lattice to form tilted squares. Along the bonds of the original
square lattice, attractive charge-charge interactions are assumed,
while on-site repulsive interactions are introduced to promote antiferromagnetism. 
The electron-Hamiltonian of such a model reads
\bea
{\cal H} & = & \sum_{i,\delta}\left[-V n_1^{i,\delta}n_2^{i,\delta}+U(n_{1\up}^{i,\delta}n_{1\down}^{i,\delta}+
n_{2\up}^{i,\delta}n_{2\down}^{i,\delta}) 
\right] \nonumber \\ 
 & & + \mbox{hopping processes} \, ,
\label{general coupling}
\eea
where the index $i$ labels the square plaquettes, while
$(i,\delta)$ denotes the four bonds extending from these plaquettes. The two
sites connected by each long bond are numbered 1 and 2 from left to right and from
bottom to top. The hopping processes can include hopping along the long and the
short bonds, as well as longer-range hopping across the square or the octagonal
plaquettes. In the large $V$, large $U$ limit, the above model reduces to one
describing the physics of spinful link-pairs, which 
reside on the long bonds of the 1/5-depleted lattice. Note that the spatial
correlations between these pairs are the same as between point-particles on
a square lattice. Since the link-pairs on different long bonds do not share
a common site, the algebra of the link-pairs on different
bonds decouples and a pseudo-spin type description of the charge sector 
becomes possible. Admittedly, this amounts to a rather radical simplification
as compared to the square lattice link-pair problem. However, the long wavelength
physics we will derive for the depleted lattice might be of a greater generality
because of the universality principle. In fact, we suspect that the complexities
discussed in the previous subsection will add only tendencies towards charge
ordering which can be to some extent discussed separately. 

\subsection{Pair-hopping and spin-spin interactions}

Since the Hamiltonian Eq.(\ref{general coupling}) should be viewed as a toy-model,
there is no reason to explicitly derive the strong-coupling description
by starting from this Hamiltonian and integrating out the states with
unpaired electrons. Instead, we simply formulate another toy-model, which
describes generic features of the dynamics of bound link-pairs on the 1/5-depleted
lattice. We include the
minimal number of processes needed to capture the physics of such a system, 
making sure that the interactions are consistent with the symmetries of the lattice.

An antiferromagnetic spin-spin interaction $J$ is assumed along 
the long bonds and a {\em ferromagnetic} interaction $J_F$ along the short
bonds.
This choice allows for an extension of the model to higher dimensions
without introducing frustration into the spin system, making it possible to 
reach the $d\rightarrow \infty$ limit and check the mean-field results
there. For $J_F\gg J$, the half-filled system becomes equivalent to an $S=2$ antiferromagnet
on a square lattice (or $S=d$ on a $d$-dimensional hypercubic lattice).
This property will be used to obtain an estimate of the quantum-corrections to 
the saddle-point results obtained in the next section.

A sublattice and an inter-sublattice hopping process are introduced, with amplitudes $t_1$ and $t_2$. Both processes
move a pair from a horizontal (vertical) bond to a nearest-neighbor vertical (horizontal) bond.
The $t_1$ process respects the spin-ordering, keeping the electrons which form
the pair on their original sublattice. The $t_2$ process moves the electrons
from one sublattice to another, thereby frustrating N\'eel order.

Including a chemical potential $\mu$, we arrive at the Hamiltonian
\bea
{\cal H} & = & \sum_i \left[\sum_{\sigma_1\sigma_2}\left\{ 
t_1\left(L_{i,\delta_x}^{\sigma_1\sigma_2\,\dagger}+L_{i,-\delta_x}^{\sigma_2\sigma_1\,
\dagger}\right)\left(L_{i,\delta_y}^{\sigma_1\sigma_2}+L_{i,-\delta_y}^{\sigma_2\sigma_1}\right) \right.\right.\nonumber \\
 & & \left.\left. + t_2\left(L_{i,\delta_x}^{\sigma_1\sigma_2\,\dagger}+L_{i,-\delta_x}^{\sigma_2\sigma_1\,
\dagger}\right)\left(L_{i,\delta_y}^{\sigma_2\sigma_1}+L_{i,-\delta_y}^{\sigma_1\sigma_2}\right) + \mbox{h.c.}\right\} \right. \nonumber \\
& & \left. -J_F\left(\vec{s}_{1 i,\delta_x}+\vec{s}_{2 i,-\delta_x}\right)\cdot\left(\vec{s}_{1 i,\delta_y}+\vec{s}_{2 i,-\delta_y}\right)\right. \nonumber \\
 & & \left. +\sum_{\delta=\delta_x,\delta_y}\left(J \vec{s}_{1i,\delta}\cdot
\vec{s}_{2i,\delta}-\mu n_{i,\delta} \right)\right]\, ,
\label{H}
\eea
where the same notation has been used as in Eq.~(\ref{general coupling}). A
projection operator $P_{i\delta}=(1-n^{i\delta}_{1\up}n^{i\delta}_{1\down})
(1-n^{i\delta}_{2\up}n^{i\delta}_{2\down})$ has been included in the
definition of the link-operators $L_{i\delta}^{\sigma_1\sigma_2\,\dagger}$, Eq.~(\ref{links}). This enforces the constraint of no double occupancy, which
is a result of the large $U$ limit in Eq.(\ref{general coupling}).

\begin{figure}[t]
\epsfxsize=0.8\hsize
\hspace{.1 \hsize}
\epsffile{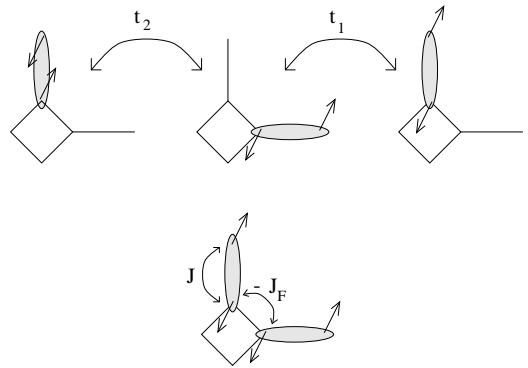}\vspace{8pt}
\caption{Hopping processes and spin-spin interactions included in the model.}
\label{f2}
\end{figure}

The Hilbert space on one long bond is spanned by five states:
unoccupied ($V$), spin-singlet ($A$) and spin-triplet ($1,0,-1$).
The operators acting on this space are $5\times 5$ matrices. Introducing the notation
\beq
\left(G_{ab}\right)_{ij}=\delta_{i,a}\delta_{j,b},
\eeq
the pair creation operators can be written as:
\bea
L_{\up\up}^{\dagger} & = & G_{1V} ,\nonumber \\
L_{\down\down}^{\dagger} & = & G_{-1V} , \nonumber \\
L_{\up\down}^{\dagger} & = & \frac{1}{\sqrt{2}}\left(G_{0V}-G_{AV}\right), \nonumber \\
L_{\down\up}^{\dagger} & = & \frac{1}{\sqrt{2}}\left(G_{0V}+G_{AV}\right). 
\label{L-en}
\eea
These operators are the equivalent of the pseudo-spins which appear in the strong-coupling
negative $U$ Hubbard model \cite{negU}. The operators $G_{\alpha V}$, $G_{V \alpha}$ and
$\frac{1}{2}(n_{\alpha}-n_V)$ form an $S=\frac{1}{2}$ spin-algebra ($\alpha=1,0,-1,A$). 
Pseudo-spins with a different spin-index $\alpha$ do not commute. In
section~\ref{ssectie SO(5) punt}, the constraint of no double 
occupancy is abandoned to allow for the construction of an $SO(5)$ symmetric version of this model. The operators
(\ref{L-en}) then become $S=1$ pseudo-spins and operators with a different
index $\alpha$ do commute in this case.

It is convenient to introduce the total spin and the N\'eel moment of a 
link-pair
\beq
\vec{S}_{i,\delta}= \vec{s}_{1 i,\delta}+\vec{s}_{2 i,\delta}
\; \; ; \; \; \vec{\tilde{S}}_{i,\delta}= \vec{s}_{1 i,\delta}-\vec{s}_{2 i,\delta} \, ,
\eeq
which are given by
\bea
S^z & = & G_{11}-G_{-1-1}\, , \nonumber \\
S^+ & = & \sqrt{2}(G_{10}+G_{0-1})\, , \nonumber \\
\tilde{S}^z & = & -G_{A0}-G_{0A}\, , \nonumber \\
\tilde{S}^+ & = & \sqrt{2}(G_{1A}-G_{A-1})\, ,
\eea
satisfying $SO(4)$ commutation relations. After  
absorbing a factor $(-1)^{i_x+i_y}\sign(\delta)$ into the singlet-state
$|A_{i,\delta}\ra$, which induces a staggering of $\vec{\St}$ and $G_{AV}$,
the Hamiltonian takes the form
\bea
 & & 
{\cal H}  =  \sum_i\sum_{\tiny \begin{array}{c}\delta_1=\pm \delta_x \\ \delta_2=\pm\delta_y\end{array}}\left[(t_1+t_2)\sum_{\alpha=1,0,-1}
\left(G_{\alpha V}^{i,\delta_1}G_{V\alpha}^{i,\delta_2}+ {\rm h.c.}\right) \right. \nonumber \\
 & & \hspace{73pt} 
+(t_1-t_2)\left(G_{AV}^{i,\delta_1}G_{VA}^{i,\delta_2}+
{\rm h.c.}\right) 
\nonumber \\
 & &  \hspace{48pt} 
 \left. -\frac{J_F}{4}\left(\vec{S}_{i,\delta_1}+\eta_i\vec{\St}_{i,\delta_1}\right)
\cdot\left(\vec{S}_{i,\delta_2}+\eta_i\vec{\St}_{i,\delta_2}\right)
\right]
\nonumber \\ & &  \hspace{.2cm} 
+\sum_{i}\sum_{\delta=\delta_x,\delta_y}\left[\frac{1}{4}J\left(1-n_V^{i,\delta}-4 n_A^{i,\delta}\right) -\mu\left(1-n_V^{i,\delta}\right)\right] \, ,
\eea
where $\eta_i=(-1)^{i_x+i_y}$ is the AF staggering factor. Note that it 
cannot be absorbed into $\vec{\St}_{i,\delta}$, since $(i,\delta_x)$ and 
$(i+1,-\delta_x)$ label the same bond.
%After  
%with $ \eta_i^{\delta}$, the Hamiltonian is rewritten on the square lattice
%formed by the bonds (dotted lines in fig.~\ref{fig het rooster}). This yields
%\bea
%{\cal H} & = & \sum_{l,\delta=\delta_x,\delta_y} \left[ %(t_1+t_2)\sum_{\alpha=1,0,-1}\left(G_{\alpha V}^{\,l}G_{V\alpha}^{\,l+\delta}+ %{\rm h.c.}\right)
%\right. \nonumber \\
% & &   \hspace{25pt} +(t_1-t_2)\left(G_{AV}^{\,l}G_{VA}^{\, l+\delta}+{\rm %h.c.}\right)
%\nonumber \\ & &
% \left.% \hspace{25pt}
%-\frac{J_F}{4}
%\left[\vec{S}_l\cdot\vec{S}_{l+\delta}+\vec{\St}_l\cdot\vec{\St}_{l+\delta}
%+(-1)^{l\cdot\delta}\left(\vec{S}_l\cdot\vec{\St}_{l+\delta}+
%\vec{\St}_l\cdot\vec{S}_{l+\delta}\right)\right] \right.\nonumber \\ 
%& & \hspace{.5cm} 
%+\sum_l\left[\frac{1}{4}J\left(1-n_V^l-4 n_A^l\right) %-\mu\left(1-n_V^l\right)\right] \, .
%\eea

\section{Mean-field analysis}

A variational Hartree-Fock procedure is used for the mean-field analysis.
In the ansatz-wavefunction, the N\'eel-vector is fixed in the $z$- and the
total spin in the $x$-direction ($\la \vec{S}\ra\cdot\la \vec{\St}\ra =0$).
The pseudo-spin degrees of freedom of the charge/phase sector are
described by an $S=\frac{1}{2}$ spin coherent state 
\bea
|\theta,\psi;\tphi^y,\chi\rangle & = &
\sin\theta {\rm e}^{-\imath \psi} |V\rangle
+ \cos\theta |\tilde{\phi}^y,\chi\rangle \, ,
\label{Cohst}
\eea
while the spin degrees of freedom of the pair
are contained in  $|\tphi^y,\chi\ra$
\bea
|\tphi^y,\chi\rangle & = & {\rm e}^{-\imath \tilde{\phi}^y\tilde{S}^y}\left(
\cos\chi |A\rangle-\sin\chi |0\rangle \right).
\label{ST}
\eea
$|\tphi^y,\chi\ra$ is just the bilayer coherent state \cite{2layer} where
the global orientation of the two-spin system has been fixed.

We list the expectation-value of a number of quantities with respect to the
variational state
\bea
 & & n=1-\langle n_V \rangle = \cos^2\theta , \nonumber \\
& &\langle S^x\rangle
 = n \sin 2\chi \sin \tilde{\phi}^y \, ; \; \la S^y\ra=\la S^z\ra=0 
, \nonumber \\
& & \langle \tilde{S}^z\rangle
  = n \sin 2\chi \cos\tilde{\phi}^y \, ; \, \la \St^x\ra=\la\St^y\ra=0
  ,\nonumber \\
& & \langle n_A \rangle
 =   n \cos^2\chi \cos^2\tilde{\phi}^y \, . \nonumber \\ & & \langle G_{\alpha V} \rangle = \sqrt{n(1-n)}
{\rm e}^{-\imath \psi} \langle \tphi^y,\chi|\alpha \rangle \, 
,  \label{verww}
\eea
where $\alpha=1,0,-1,A$ labels the four spin-states.
The role which the various parameters play can
be determined from this list:$\theta$ fixes the pair-density;
$\tilde{\phi}^y$ determines the relative magnitude of $\la \vec{S}\ra$ 
and $\la \vec{\tilde{S}}\ra$, while their total magnitude  is 
fixed by $\chi$;
$\psi$ represents the phase which orders in the superconducting state.

The variational energy is given by
\beq
E_{\rm var.}(\{\theta,\psi;\tphi^y,\chi\}_{i,\delta})
= \langle \{i,\delta\}|
{\cal H}| \{i,\delta\}\rangle \; ,
\eeq
where
\beq
|\{i,\delta \}\rangle= \prod_{i,\delta}|\theta,\psi;\tphi^y,\chi\rangle_{i,\delta} .
\eeq

\begin{figure}[h]
\vspace{-12cm}
\epsfxsize= \hsize
%\hspace{.0\hsize}
\epsffile{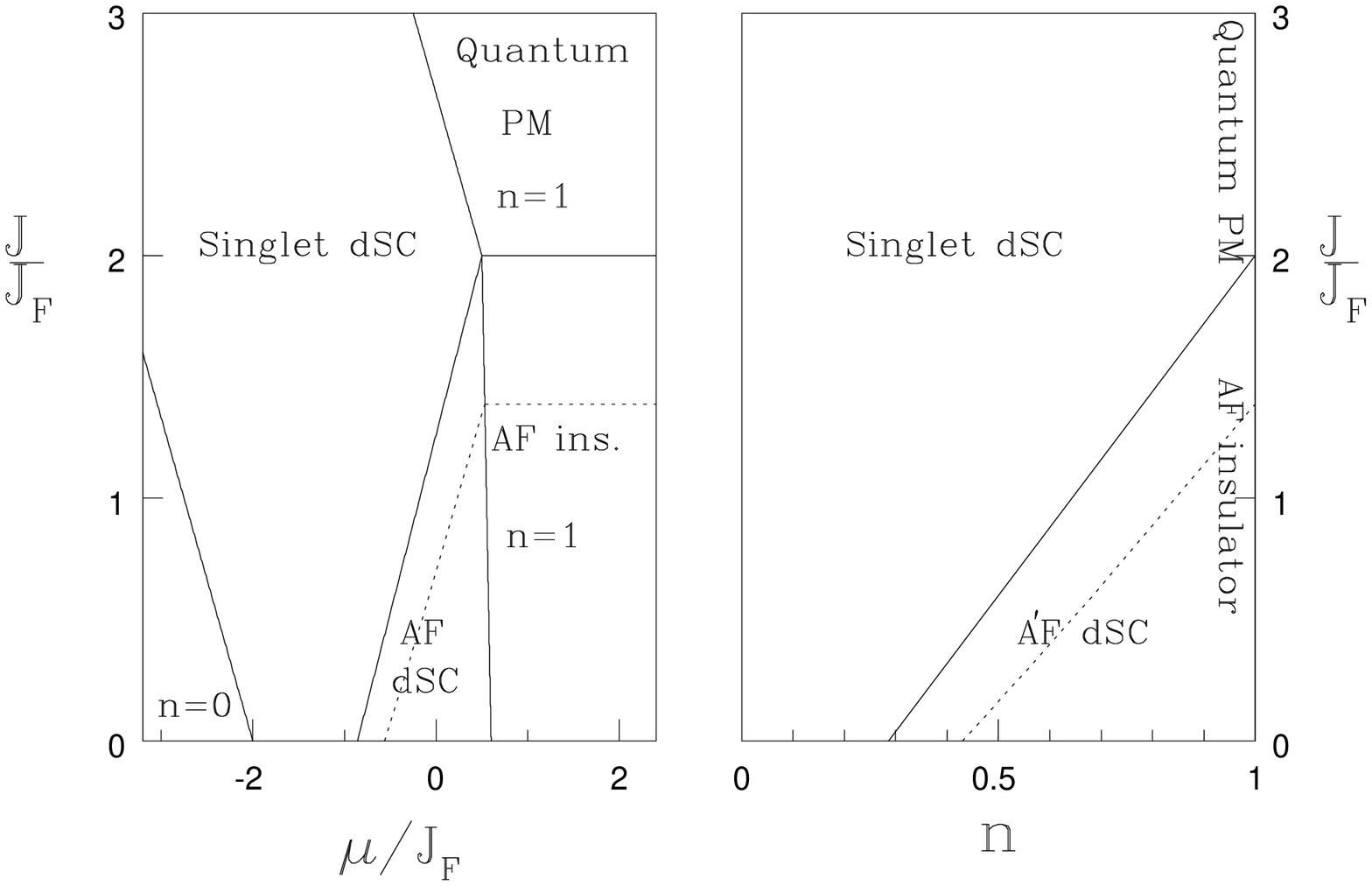}
%\vspace{-2cm}
%\caption{Mean field phase diagram of $J$ versus $\mu$ and $n$, for
% $t_1>t_1^*$ \,  ($t_1=0.4 J_F$, $t_2=-0.1 J_F$).  The dotted line indicates the quantum
%corrections if transversal spin-fluctuations are taken into account.}
%\label{f3}
%\end{figure}
%\begin{figure}[t]
\vspace{-2.5cm}
\epsfxsize=\hsize
%\epsfysize=0.8 \hsize
%\hspace{.1 \hsize}
\epsffile{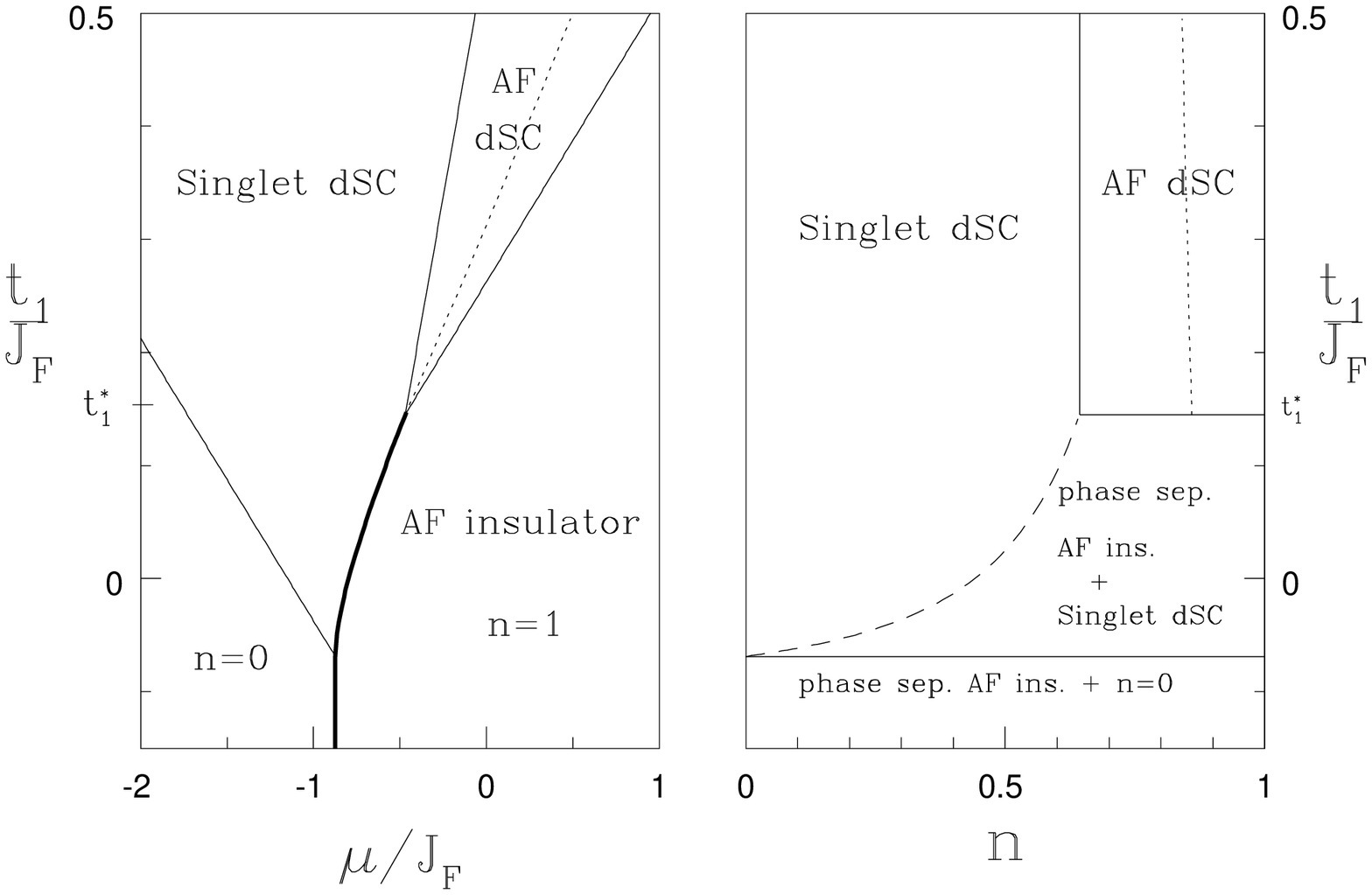}
%\vspace{-2cm}
%\caption{Mean field phase diagram of $t_1$ versus $\mu$ and $n$, for
% $J<2 J_F$ and $t_2<J/8$ ($J=J_F$, $t_2=-0.1 J_F$).  The dotted line indicates the quantum
%corrections if transversal spin-fluctuations are taken into account.
%Bold lines represent first-order phase transitions.}
%\label{f4}
%\end{figure}
%\begin{figure}[t]
\vspace{-2.5cm}
\epsfxsize=\hsize
%\hspace{.1 \hsize}
\epsffile{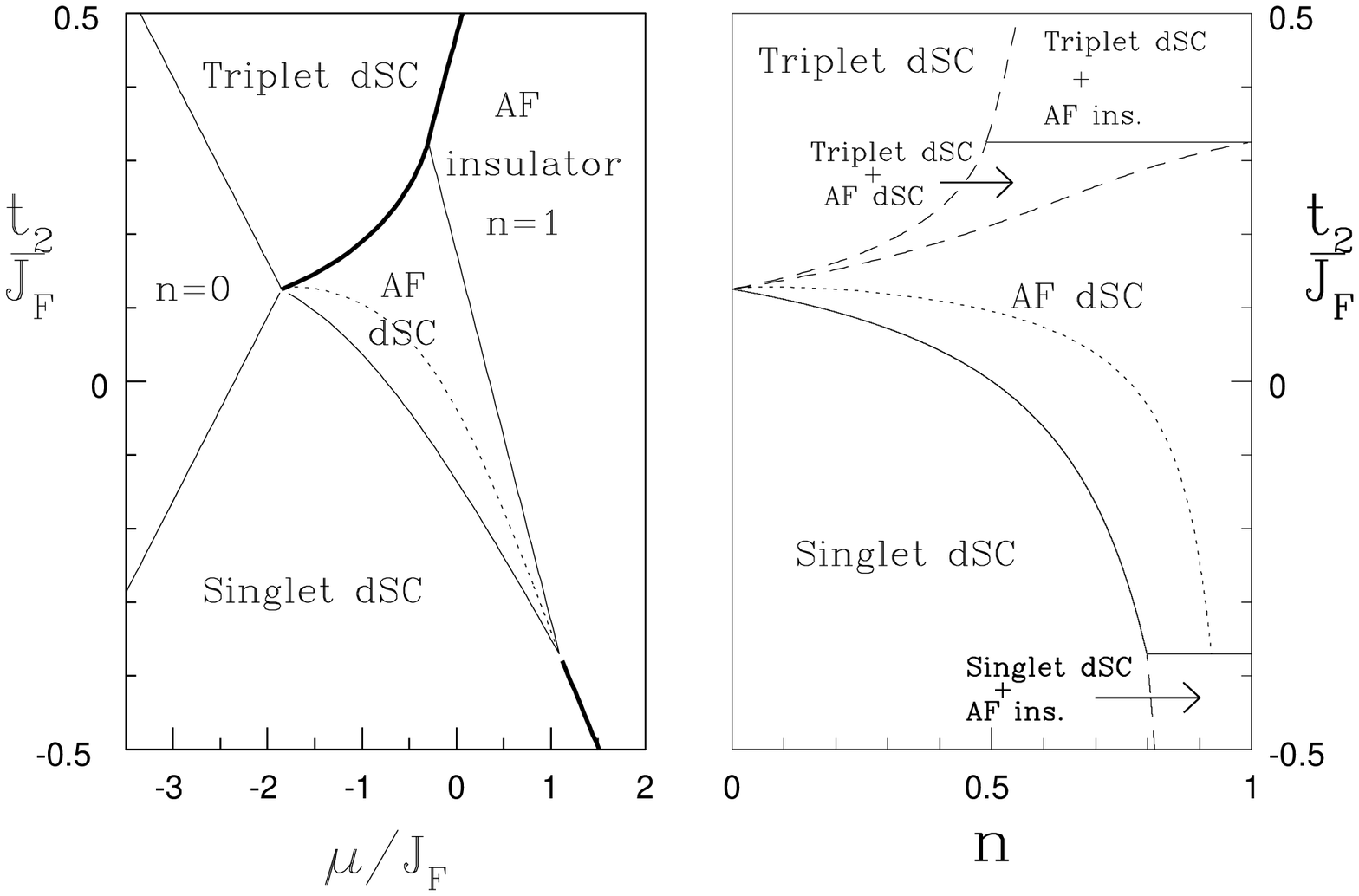}
\vspace{-1cm}
\caption{Mean field phase diagram of $J$ versus $\mu$ and $n$, for
 $t_1>t_1^*$ \,  ($t_1=0.4 J_F$, $t_2=-0.1 J_F$).  Bold lines indicate
first order transitions. At the dotted line, transversal quantum
spin-fluctuations destroy the antiferromagnetic order.}
 \label{f3}\vspace{8pt}
\caption{Mean field phase diagram of $t_1$ versus $\mu$ and $n$, for
 $J<2 J_F$ and $t_2<J/8$ ($J=J_F$, $t_2=-0.1 J_F$).}
\label{f4}\vspace{8pt}
\caption{Mean field phase diagram of $t_2$ versus $\mu$ and $n$, for $J<2 J_F$ and
 $t_1>(4J_F^2+J^2)/32J_F$ \, ($t_1=0.4 J_F$, $J=J_F$).}
\label{f5}\vspace{8pt}
\end{figure}

In the mean-field analysis, it is assumed that the staggered local
magnetization and the charge density are uniform. The phase $\psi_l$ is 
allowed to have a different value on horizontal ($\psi^{\rm H}$) 
and vertical bonds ($\psi^{\rm V}$). We then 
arrive at the following mean-field energy
\bea
& & E_{\rm MF}  = 
N\left( \sin^2 2\theta (t_1+t_2-2 t_2 \cos^2\chi\cos^2\tilde{\phi}^y)\times \right. \nonumber \\
 &  & \left.\times  \cos(\psi^{\rm H}-\psi^{\rm V}) -\frac{1}{2}J_F \cos^4\theta \sin^2 2\chi 
\right. \nonumber \\
 & & \left.
+\frac{1}{4}J \cos^2\theta(1-4 \cos^2\chi\cos^2\tilde{\phi}^y)- \mu \cos^2\theta\right),
\label{Emf}
\eea
where $N$ denotes the number of long bonds. Minimizing Eq.~(\ref{Emf}), a variety of 
mean-field groundstates is obtained as a function of
the various parameters. The results 
are summarized in figures \ref{f3}-\ref{f5} and in table 1. 
We focus here on the case
$t_1>0$, for which the superconducting state is typically of d-wave type 
($\psi^{\rm H}-\psi^{\rm V}=\pi$). The same phase-diagram results for $t_1\rightarrow -t_1$ and $t_2\rightarrow -t_2$, but
with s-wave instead of d-wave phase-order. For simplicity,
$J$, $t_1$, $t_2$ and $\mu$
are expressed in units of $J_F$ from here on.

At half-filling, the physics is determined by the competition between the
antiferromagnetic and the ferromagnetic spin-spin interaction. While the 
first promotes singlet-formation along the horizontal and vertical bonds,
the second favors large local magnetic moments. $J$ therefore tunes
the singlet density in the
groundstate at half-filling. For $J\ll 1$, the system has full N\'eel order with
$\la n_A\ra =\frac{1}{2}$, $|\la \vec{\tilde{S}} \ra |=1$. The singlet density increases linearly with $J$ up to $\la n_A\ra=1$ at $J=2$, where the staggered
magnetization vanishes in a second order transition to a quantum paramagnet 
phase (Fig.~\ref{f3}).  

For densities smaller than one, the two hopping processes begin to play a role.
Since the case of a uniform charge-distribution is considered and since
all electrons are paired in the strong-coupling limit, all variational
states with a non-integer electron-density exhibit superconductivity.
The superconducting order-parameter depends on the electron-density
as $|\Delta|\sim \sqrt{n(1-n)}$, see Eq.(\ref{verww}).

The value of the hopping amplitude $t_1$ determines the nature
of the transition from the 
antiferromagnetic insulator at half-filling to the singlet superconductor
at lower densities. For small $t_1$, this transition
is first order as a function of $\mu$, giving rise to a region of antiferromagnet/
superconductor phase separation in the $t_1$-$n$ phase-diagram (Fig.~\ref{f4}). At
$t_1=t_1^*=\frac{1}{8}+2 t_2^2$, the first order line splits into two second order
lines. A region opens up in which the system has both antiferromagnetic
spin order and superconductivity. In this antiferromagnetic superconductor
(AFSC) phase, the electrons which carry the superconducting order-parameter are
at the same time responsible for the antiferromagnetism. This state is most
easily visualized by thinking of a small density of nearest-neighbor
hole-pairs being doped into a half-filling antiferromagnet. If these hole-pairs
are most mobile along the diagonals of the square lattice, where their
movement does not disturb the AF spin order, they can  
delocalize and in that way give rise to superconductivity {\em without}
at the same time destroying the antiferromagnetic order. The condition
that diagonal pair-hopping has to dominate to get an AFSC-phase on the
square lattice is reflected by the condition $t_1>t_1^*$ for the present model.

There are three ways in which the spin order-parameter in the 
AFSC phase is suppressed through the doping with hole-pairs. The simplest one corresponds with the dilution 
of  the antiferromagnet by the removal of spins. More interestingly, the inter-pair spin-spin interaction
$J_F$ scales with the pair-density squared, while the intra-pair 
spin-spin interaction $J$ scales linearly with $n$. As a result, the 
ferromagnetic interaction is suppressed by a factor $n$ relative to $J$,
pushing the ratio $J/J_F$ closer to its critical value and reducing the magnetic
moment per pair. Finally, the hopping process 
$t_2$ frustrates the N\'eel order,
provided that $\sign(t_2)=-\sign(t_1)$ (the other case is discussed below). 
The increase of the singlet density per pair due to the last two processes
results in a transition to the singlet superconductor at $n=n_c=\frac{J-8 t_2}{2-8 t_2}$.

Since the $t_2$ process amounts to a $t_1$-type hop with an additional interchange
of the two electrons forming the pair, it picks up a minus-sign when acting on a
pair in the antisymmetric spin-singlet state. 
Suppose that $t_1$ and $t_2$ have 
the same sign. A singlet-pair, through the $t_2$-process,
then frustrates the phase-ordering as favored by the $t_1$-hop. To reduce this 
frustration, the singlet
content of the pairs is suppressed as $t_2$ is increased, enhancing the
spin-ordering in the AFSC phase. Eventually, a first order transition
occurs to a ferromagnetically ordered triplet superconductor phase, where
the singlet-density is reduced to zero (Fig.\ref{f5}).
%hierwashet

If $t_1$ and $t_2$ have opposite sign, the {\em triplet} component is suppressed
through the same process and the $t_2$-hop reduces the spin-order in the AFSC phase.
Note that $t_2$ causes a positive shift of the critical value $t_1^*$ regardless
of its sign. This reflects the fact that on-sublattice hopping must dominate
in order for an AFSC phase to occur ($t_1>t_1^*$ implies $t_1\geq |t_2|$,
where the equal sign occurs for $|t_2|=\frac{1}{4}$).

It should be verified that the saddle-point solution becomes 
exact in the limit $d\rightarrow \infty$. To reach this limit, the model has to 
be formulated in arbitrary dimension. The d-wave phase order then posses a problem, 
since it cannot be generalized to dimensions higher than 2. However, the
Hamiltonian of the 2d system is invariant under a 
simultaneous sign-change of $t_1$, $t_2$ and $G_{\alpha V}^{i,\pm\delta_x}$,
which implies that d- and s-wave order are equivalent for the 2d model
\footnote{This no longer holds if longer-range hopping processes are included.}. We therefore
flip the sign of $t_1$ and $t_2$ and study the $d\rightarrow \infty$ limit for the 
s-wave ordered state. In order to keep the energy finite,  $J_F$, $t_1$ and $t_2$ are scaled
with $\frac{1}{d}$ while taking the limit. The variation of the energy is found to be
of order $\frac{1}{d}$ for the saddle-point solution. Since it
vanishes at large $d$, the mean-field groundstate indeed becomes 
an eigenstate of the system in this limit. 

\vspace{6pt}
%\begin{table}[b]
\begin{centering}
\begin{tabular}{|l|c|c|c|}\hline
phase & $n$ & $\cos 2\chi$ & $\cos \tilde{\phi}^y$ \\ \hline
 & & & \\
Spin-liquid & 1 & 1 & 1 \\ %\hline
N\'eel dSC & $\frac{-J-4\mu-16 t_1-8 J t_2+64 t_2^2}{4(1-8t_1+16t_2^2)}$
& $\frac{J+8 t_2(n-1)}{2 n} $
& 1 \\ %\hline
Singlet dSC & $\frac{4 \mu+16 (t_1-t_2)+3J}{32(t_1-t_2)}$ & 1 & 1 \\ %\hline
Triplet dSC & $\frac{J-4\mu-16(t_1+t_2)}{4-32(t_1+t_2)}$ & 0 & 0  \\ %\hline
AF & 1 & $\frac{J}{2}$ & 1 \\ %\hline
 & & & \\ \hline
\end{tabular}\\ \vspace{8pt}
%\caption{Mean-field results for the various phases.}
TABLE 1: Mean-field results for the various phases.\\
\end{centering}
%\end{table}

\section{Transversal spin fluctuations}
\label{ssectie trans fluc}

In the AFSC phase, both the $U(1)$ phase- and the $SU(2)$ spin-symmetry are spontaneously broken. As a result, the system has two spin-wave modes 
and one phase Goldstone mode. These gapless modes dominate its 
long-wavelength physics. Since they decouple, 
the phase and spin degrees of freedom of the system 
may be treated separately at sufficiently large lengthscales.

The physics of the phase sector is equivalent to that of an XY-spin model in an external
magnetic field, which has a dynamical critical exponent $z=2$. The $T=0$ system
is therefore effectively at its upper critical dimension $d=2+z=4$.
Because of the high effective dimensionality, phase-fluctuations only give
small correction to the zero-temperature mean-field 
results for the insulator-superconductor transition \cite{Anne}.

The long-wavelength behavior of the spin-sector in the AFSC phase is
characterized by a critical exponent $z=1$. Hence, at $T=0$, the spin-sector 
lives
effectively in three dimensions and fluctuation-effects
can be significant. The long-wavelength spin physics is described by
an effective non-linear sigma model \cite{Haldane,CHN}. This model contains
one coupling constant, $g_0$, which is a measure of the quantum fluctuations
in the system. At a critical value of $g_0$, the spin-system
undergoes a quantum phase transition
from a N\'eel ordered state to a quantum paramagnet.
For the present model, $g_0$ is expected to 
diverge at the mean-field
transitions to the singlet superconductor and the paramagnetic insulator 
\cite{2layer}. Transversal fluctuations are therefore expected to
significantly  reduce the region in the phase-diagram where AF order is
stable. 

The coupling constant of the effective non-linear sigma model 
depends on the bare values of the
spin stiffness and the perpendicular susceptibility,
which are properties of the microscopic model. 
The mean-field expressions for these quantities can serve as an estimate for
their bare value \cite{2layer}.
These expressions are derived below for the present model.

\begin{figure}[t]
\epsfxsize=.8\hsize
\hspace{.1\hsize}
\epsffile{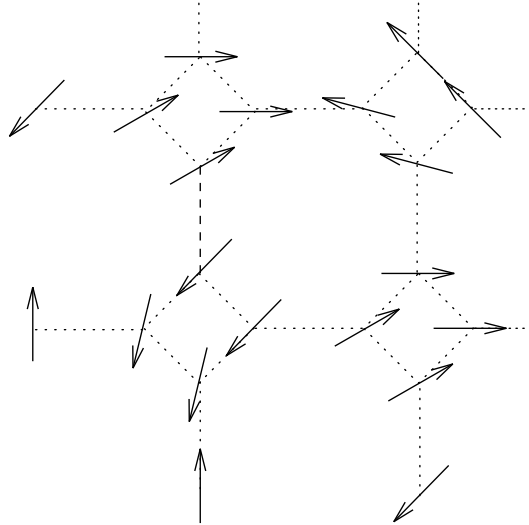}\vspace{8pt}
\caption{A spin-configuration with a twist along the $x+y$-direction.}
\label{fig de twist}
\end{figure}

We define the perpendicular susceptibility as the induced magnetization
per square plaquette (containing four spins) by a vanishing magnetic field applied
perpendicular to the direction of antiferromagnetic ordering.
It is calculated by adding a magnetic field term to the mean-field energy Eq.~(\ref{Emf}),
\beq
H \sum_{i,\delta=\delta_x,\delta_y} \la S^x_{i,\delta}\ra =N H \cos^2\theta 
\sin 2\chi \sin \tilde{\phi}^y \, \label{Emagn}
\eeq  
and subsequently minimizing the energy. This yields
\bea
\chi_{\perp} & = & \lim_{H\rightarrow 0} \frac{2\la S^x\ra}{H} = \frac{2(1-\cos 2\chi) n}{J-8 t_2(1-n)} \nonumber \\
& \Rightarrow & \left\{\begin{array}{rcl}\chi_{\perp}^{\rm AF} & = & \frac{2-J}{J} \\
 & & \\
 \chi_{\perp}^{\rm AFSC}   & = & \frac{n(2-8 t_2)-J+8 t_2}{J-8 t_2(1-n)} 
\end{array}\right.
\label{chi pair}
\eea
The susceptibility vanishes at the transitions to the quantum paramagnet and the singlet superconductor phase.
It has a divergence at $n=1-\frac{J}{8t_2}$, which is interrupted by the first order transition to the triplet superconductor phase (Fig.~\ref{f5}) (the
line where $\chi_{\perp}$ diverges and the first-order line approach each other for small $n$). 

For $n=1$ and $J_F\gg J$, the four 
spins around each square plaquette lock into a symmetric state. The 
spin-operators can then be replaced by $\frac{1}{4}$ 
times the total spin-operator on the plaquette. The resulting
Hamiltonian describes an $S=2$ antiferromagnet on a square
lattice, with a spin-spin coupling  $J_{\rm eff}=\frac{1}{16}J$.
Such a system has a mean-field perpendicular susceptibility 
$\chi_{\perp}=1/8J_{\rm eff}=2/J$ \cite{Haldane}
(where the lattice-spacing, in this case the distance between neighboring
square plaquettes, is set to one). The above result for the susceptibility indeed 
reduces to this expression for $n=1$ and $J\ll 1$.

To determine the spin stiffness, the configuration shown in Fig.~\ref{fig de twist} is 
considered. It has a slow twist in the spin order parameter
along the $x+y$-direction. The stiffness gives the lowest-order correction
to the groundstate energy due to this twist \cite{Igar}.

At each antiferromagnetic bond along the direction of the twist,
 the spins have been rotated over an angle $\alpha\, \delta \phi$ in
the XZ-plane, at each ferromagnetic bond over an angle $(1-\alpha)\, \delta \phi$. 
This configuration
is described by the variational state Eq.~(\ref{Cohst}), where the
spin-part is given by
\beq
{\rm e}^{\imath\,  l \,\delta\phi \,S^y} \left|\tphi^y=\frac{1}{2}\alpha\delta\phi,\chi=\chi_{MF}\right\ra \, ,
\eeq
with the index $l$ labelling the bonds along the twist.
 
The antiferromagnetic interaction energy of this state is given by
(compare with last
line in Eq.(\ref{Emf}))
\bea
& & \frac{1}{4}J N \cos^2\theta\left(1-4 \cos^2\chi\cos^2(\frac{1}{2}\alpha\delta\phi)\right)\simeq \nonumber \\ 
 & & E_{AF}(\delta\phi=0)+\frac{1}{4}J N \alpha^2 \delta\phi^2 \cos^2\theta\cos^2\chi \,,
\eea
while the ferromagnetic energy is simply reduced by a factor $\cos (1-\alpha)\,\delta\phi$ per twisted ferromagnetic bond. 
The phase-ordering energy also contributes to the spin stiffness. Along the twist, we have
(d-wave order)
\bea
& & \sum_{\alpha=1,0,-1}\la G_{\alpha V}^l\ra\la G_{\alpha V}^{l+1}\ra = \nonumber \\
 & & -\sum_{\alpha=1,0,-1}\frac{1}{4}\sin^22\theta\la\vec{\Omega}|{\rm e}^{-\imath\, l \,\delta\phi \,S^y}
|\alpha\ra\la\alpha|{\rm e}^{\imath\, (l+1) \,\delta\phi \,S^y}|\vec{\Omega}\ra
= \nonumber \\
 &  &  -\frac{1}{4}\sin^22\theta\la \vec{\Omega}|\left(\Unity-|A\ra\la A|\right){\rm e}^{\imath\, \delta\phi\, S^y}|\vec{\Omega}\ra \simeq \nonumber
\\
 & &  -\la G^l_{AV}\ra\la G^{l+1}_{VA}\ra -\frac{1}{4}\sin^22\theta\left(1-\frac{1}{2}\delta\phi^2 
\underbrace{\la \vec{\Omega}|S^{y\,2}|\vec{\Omega}\ra}_{\sin^2\chi} \right)\, , \nonumber \\
& & \la G^l_{AV}\ra\la G^{l+1}_{VA}\ra = \nonumber \\
 &  &  -\frac{1}{4}\sin^22\theta \cos^2\chi\cos^2\left(
\frac{1}{2}\alpha^2\,\delta\phi^2\right) \simeq\nonumber \\
 & &   -\frac{1}{4}\sin^22\theta \cos^2\chi \left(1-\frac{1}{4}\alpha^2\delta\phi^2\right) \, .
\eea
Taking these contributions together, the energy-increase due to the twist in 
the spin order-parameter is found to be
\bea
\Delta E & =  & \frac{1}{8}N\,\delta\phi^2\left(2 \sin^22\theta\left[(t_1+t_2) 
\sin^2\chi-t_2 \alpha^2\cos^2\chi\right] 
  \right. \nonumber \\
 & & \left. + \sin^22\chi\cos^4\theta(1-\alpha)^2 
   +2 J \alpha^2 \cos^2\theta \cos^2\chi \right)
% \nonumber \\
% & \equiv & \frac{N}{2} \delta\phi^2\rho_s \, .
\label{rhos pairs}
\eea
 
The distribution of the total twist over the two types of bonds is 
obtained by minimizing this energy with respect to $\alpha$, which
yields
\beq
\alpha_0=\frac{ n(1-\cos 2\chi)}{n(1-\cos 2\chi)+J-4 t_2(1-n)} \, .
\eeq  
For $J_F$ much larger than $J$ and $t_2$ (or just $J$ for $n=1$), the
twist is entirely localized on the antiferromagnetic bonds, as expected.
At the transition to the
spin disordered phases, it is localized on the ferromagnetic bonds.

The stiffness now follows from 
\beq
\Delta E(\alpha_0)=\frac{N}{2}\delta\phi^2\rho_s \,.
\label{rhodef}
\eeq
For the half-filling antiferromagnet, we obtain
\beq
\rho_s^{\rm AF} = \frac{1}{8}J(2-J)\, ,
\eeq
while for the AFSC phase the stiffness is given by
\bea
& & 
\rho_s^{\rm AFSC} = \frac{n(2-8 t_2)-J+8 t_2}{8J+16n} \times
\label{rh AFSC} \\
 & & \times
\left[2J n+J^2+4(1-n)\left(J(t_1-2 t_2)+2 t_1 n + 8 t_2^2(1-n)\right)\right] \, .
\nonumber
\eea
 Like the susceptibility, the stiffness vanishes at the 
transition to a spin-disordered phase. It reduces to the $S$$=$$2$, $J_{\rm eff}=\frac{1}{16}J$-form for $J\gg J_F$ at half-filling. 

The bare coupling constant of the non-linear sigma model is given by
$g_0=(\rho_s\chi_{\perp})^{-\frac{1}{2}}$. As expected, it diverges at the
transitions to the singlet superconductor and quantum paramagnet, since
both the susceptibility and the stiffness vanish in these phases.
In order to obtain a more precise estimate of $g_0$, its value is
shifted by a constant factor such that it agrees with the
result for the $S$$=$$2$ antiferromagnet at $n=1$, $J_F\gg J$. The bare
coupling for the square lattice $S$$=$$2$ antiferromagnet can be determined
from spin-wave results for the renormalized spin-wave 
velocity 
and perpendicular susceptibility, using the one-loop expression \cite{CHN}
\beq
\frac{g_0}{4\pi}=\frac{1}{1+4\pi\chi_{\perp} c/\hbar\Lambda}\, ,
\eeq
where $\Lambda=2\sqrt{\pi}/a$, with $a$ the lattice spacing.
Using the spin-wave results of Igarashi \cite{Igar}, we obtain
\beq
g_0^{S=2}\simeq 3.85 \, .
\eeq

For the half-filling antiferromagnet, the bare coupling constant is given by
\beq
g_0=g_0^{S=2}\frac{2}{2-J}\, ,
\eeq
while we find for the AFSC phase,
\bea
 & & g_0  =  g_0^{S=2}\,\frac{2\sqrt{(2n+J)(J-8 t_2(1-n))}}{n(2-8t_2)-J+8 t_2
} \times
 \\
 &  & \times  \frac{1}{
\sqrt{2nJ +J^2 +4 (1-n)[J(t_1-2t_2) + 2 t_1 n
+ 8 t_2^2(1-n)] } } \, .\nonumber
\eea

The order-disorder transition at $g_0=g_c=4\pi$ is indicated by
a dotted line in the mean-field phase diagrams, Fig.s \ref{f3}-\ref{f5}. 
It is found that transversal spin fluctuations 
significantly reduce the parameter-range over which the 
N\'eel-ordered phases are stable, without changing the topology of the
zero-temperature phase diagram.
%In the region between the mean-field and the quantum-disordering
%line, the system still has finite local magnetic moments, but these are
%fluctuating and do not form static antiferromagnetic order. Given the
%evidence from neutron-scattering, $\mu$SR and NMR
%for at least slow spin-dynamics 
%and possibly static (incommensurate) AF order at low temperatures
%in the underdoped high-$T_c$ superconductors,
%it seems reasonable to identify these systems in the framework of the present
%model with the parameter-values in the neighborhood of the quantum-critical
%transition between the AFSC and the singlet superconductor phase. 

At non-zero but low temperatures, the quantum nonlinear sigma model 
predicts $z=1$ quantum
critical behavior in a parameter region around the AFSC to SC transition
line \cite{CHN}. The 
width of this region grows as $|g_0-4\pi|\sim T^{-\nu}$, with $\nu=0.7$ the correlation-length critical exponent of the 3d
Heisenberg model. This type of finite temperature behavior, where 
temperature becomes the only energy-scale in the system, has been
reported for the underdoped cuprates by a number of authors
\cite{Aeppli,B P,Sachdev}. 

Finally, we note that in the present model the superconductivity
onset-temperature is completely determined by phase-fluctuations. This is
a trivial consequence of the fact that we focussed on the strong-pairing 
limit. Nevertheless, it is consistent with recent analyses of the dependence
of $T_c$ on the zero-temperature phase-stiffness and on the number
of closely-spaced layers in the superconductor material, which point
to a dominant role of finite-temperature phase-fluctuations in determining
$T_c$ \cite{classphase}.

%behavior opens up around these lines. The width of this region in terms of the microscopic
%parameters is $|g_0-4\pi|^{\nu}$, with $\nu=0.7$ the exponent of the 3d Heisenberg model  \cite{CHN}. 
%The quantum critical region is characterized by `$\omega/T$-scaling' (temperature sets the energy-scale 
%in the system). This behavior has been observed by a number of groups for the underdoped high-$T_c$
%superconductor \cite{Aeppli,BP}. As was pointed out in the introduction.... 
% This agrees with the observations of Aeppli {\em et al.}
%of quantum-criticality in the spin sector of underdoped La$_{2-x}$Sr$_x$CuO$_4$. 
%It was argued in \cite{Janstuk} that the explanation of this behavior
%in terms of the nearby charge- and spin-ordered stripe phase requires the
%existence of a coexistence phase. From this analysis, we find that such a %coexistence
%phase can in principle exist, and that it indeed gives rise to a %superconducting state with
%quantum-critical behavior in the spin-sector.

\section{SO(5) symmetric point}
\label{ssectie SO(5) punt}

The AFSC phase has an interesting property. Let us consider the $SO(5)$
superspin-vector \cite{Zhangscience} for this model.
\bea
\vec{N}_{\rm P}& = & \left(\frac{1}{2}(G_{AV}+G_{VA}),\frac{1}{2}\vec{\tilde{S}},\frac{1}{2 \imath} (G_{AV}-G_{VA})\right).
\label{superspin}
\eea
The label `P' indicates that $\vec{N}_{\rm P}$ is defined in the projected Hilbertspace, where double site-occupancy is forbidden. The mean-field expectation value of $\vec{N}_{\rm P}$ 
satisfies
\beq
\left.\frac{\partial \la \vec{N}_{\rm P} \ra^2}{\partial n}\right|_{t_2=-\frac{1}{4}} =0.
\eeq
Hence, at the mean-field level and for this particular choice of $t_2$, the AFSC phase can be characterized by an $SO(5)$ order-parameter which has components
both in the superconducting and in the antiferromagnetic subspace, and which
is rotated from the AF to the SC direction as the hole-density is increased. 
As one approaches the tricritical point, the AFSC states with different
$n$ become degenerate
(Fig.\ref{f3}) and the mean-field state becomes invariant under rotations of $\vec{N}_{\rm P}$.
For $t_2=-\frac{1}{4}$ the tricritical-critical point is located at $t_1=t^*=\frac{1}{4}$,
$\mu=\mu^*=\frac{J}{4}$.

It should perhaps come as no surprise that we find a `mean-field $SO(5)$ 
symmetry'
for this model. The special lattice used here has two orbitals
per unit cell, which 
seems to be one of the requirements for constructing an $SO(5)$
symmetric model with short-range interactions \cite{SCso5}. This 
can be understood from the fact that the minimum number of sites 
required for the electron Hilbert-space in which an $SO(5)$ representation can
be constructed is two (since the $\pi$-operators are spin-1, charge 2 objects). Two-leg spin ladders have a natural two-site unit,
the rung, on which the $SO(5)$ order parameter can be defined. The lattice 
used here also has such a unit: the long bond. To formulate
a short-range $SO(5)$ model on the square lattice, one either has to break the lattice-symmetry,
or to involve a certain amount of 
coarse-graining, which means that the resulting $SO(5)$-description is 
effective rather than microscopic. 

In the following, 
an exact $SO(5)$ symmetric point is derived for the present model. The procedure
used is similar to that for the $SO(5)$ symmetric ladder \cite{SO5ladder}.
At the mean-field $SO(5)$-point, the Hamiltonian is given by
\bea
{\cal H} & = & {\cal H}_0 +{\cal H}_1 ,
\eea
where
\bea 
{\cal H}_0  & = & -\sum_{i}\sum_{\tiny \begin{array}{c}\delta_1=\pm \delta_x \\ \delta_2=\pm\delta_y\end{array}} \vec{N}_{\rm P}^{i,\delta_1}\cdot
\vec{N}_{\rm P}^{i,\delta_2}-J\sum_{i}\sum_{\delta=\delta_x,\delta_y}n_{A}^{i,\delta} \, , \label{H0} \\
{\cal H}_1  & = & \sum_{i}\sum_{\tiny \begin{array}{c}\delta_1=\pm \delta_x \\ \delta_2=\pm\delta_y\end{array}}
\frac{1}{4}\left[\vec{\tilde{S}}_{i,\delta_1}\cdot\vec{\St}_{i,\delta_2}
\right.\nonumber \\
 & & \left. +\eta_i\left(\vec{S}_{i,\delta_1}\cdot\vec{\St}_{i,\delta_2}
+ \vec{\St}_{i,\delta_1}\cdot\vec{S}_{i,\delta_2}\right)\right]  \, ,
\label{H1}
\eea
absorbing a d-wave staggering into the $|V_{i,\delta}\ra$-state.

The second term, ${\cal H}_1$, is manifestly not invariant under rotations
of $\vec{N}_P$. After decoupling the operators on different bonds with respect
to the order-parameters for superconductivity and antiferromagnetism, 
this term vanishes and therefore the
symmetry breaking does not show up at the mean-field level. 
As Eder {\em et al.}
pointed out \cite{Eder}, the first and the fourth component of $\vec{N_{\rm P}}$
are rotated into each other by transforming the zero-magnetization triplet
state $|0\ra$ into the hole-pair state $|V\ra$. This transformation leaves the
singlet density $n_A$ invariant. Since one may assume in mean-field
that all components of $\la \vec{N}_{\rm P}\ra$ vanish except the first and the
fourth (spontaneous symmetry breaking selects a preferred direction in the spin- and phase sector) the decoupled mean-field Hamiltonian 
is invariant under this transformation. This implies that the $d\rightarrow \infty$ $SO(5)$-symmetry is not only present in the zero-temperature
groundstate, but also at finite temperatures, where higher
energy-levels are thermally occupied. 
  
As a first step towards an $SO(5)$-symmetric Hamiltonian,
${\cal H}_1$ is subtracted from ${\cal H}$. This introduces
second- and third-neighbor spin-spin interactions into the model. 

The second term in ${\cal H}_0$ is an $SO(5)$-invariant 
(this is discussed below).
The first term is invariant under rotations of $\vec{N}_P$ , but this
does not imply that it is $SO(5)$ symmetric. There is no 
representation of the $SO(5)$ algebra on the projected Hilbertspace under which
$\vec{N}_{\rm P}$ transforms as a vector. The rotation-symmetry is therefore broken at
the quantum level. In a recent article \cite{pSO5}, Zhang {\em et al.} 
show that mean-field $SO(5)$ symmetry always remains when a projection to
the lower Hubbard band is applied to a system with full $SO(5)$ symmetry. Here
we work backwards: mean-field $SO(5)$-symmetry being established, we deduce a model
with full $SO(5)$ symmetry by lifting the constraint of no double site-occupancy.

% We will call ${\cal H}_0$ {\em classically}$SO(5)$symmetric, using the term
%SO(5) somewhat loosely to signify rotations of the classical 
%5-vector $\la \vec{N}_{\rm P}^{l}\ra $.

The basis of the single-bond Hilbert-space is extended
to include the doubly occupied state $|D\ra$. It now
consists of one $SO(5)$ singlet ($A$) and one $SO(5)$ quintet (spin-triplet, $D$ and 
$V$). The details of this representation of the
$SO(5)$ algebra are briefly discussed in appendix~\ref{App1}. We introduce an 
on-site repulsion $U\sum_{i\delta}n_{D\,i\delta}$. 
The general Hamiltonian on the unprojected Hilbert space is given by
\bea
& & {\cal H} = \sum_{i}\sum_{\tiny \begin{array}{c}\delta_1=\pm \delta_x \\ \delta_2=\pm\delta_y\end{array}}\left[ 4(t_1+t_2)\sum_{\alpha}\vec{\pi}^{\alpha}_{i,\delta_1}\cdot\vec{\pi}^{\alpha}_{i,\delta_2} \right. \nonumber \\
 \nonumber \\
%& & -\sum_{i}\sum_{\tiny \begin{array}{c}\delta_1=\pm \delta_x \\ \delta_2=\pm\delta_y\end{array}}\left[ 4(t_1+t_2)\left(\Re\vec{\pi}_{i,\delta_1}\cdot\Re\vec{\pi}_{i,\delta_2}
%+ \Im\vec{\pi}_{i,\delta_1}\cdot\Im\vec{\pi}_{i,\delta_2}\right) \right. \nonumber \\
 & & \left.
+(t_1-t_2)\vec{\Delta}_{i,\delta_1}\cdot\vec{\Delta}_{i,\delta_2} 
\right. \nonumber \\ & & \left.
+\frac{1}{4}\left(\vec{S}_{i,\delta_1}+\eta_i\vec{\St}_{i,\delta_1}\right)
\cdot \left(\vec{S}_{i,\delta_2}+\eta_i\vec{\St}_{i,\delta_2}\right) \right] \label{Hup} \\
% & & 
%+\sum_{i,\delta=\delta_x,\delta_y}\left[ \frac{J}{4}(1-4 n_A^{\, i,\delta})+\left(U-\frac{J}{4}-\mu\right)n_D^{\, i,\delta}
% + \left(\mu-\frac{J}{4}\right)n_V^{\, i,\delta} \right] \, ,\nonumber\\
% & & +\frac{1}{4} \sum_{i,\delta=\delta_x,\delta_y}\left[ J-4 J n_A^{\, i,\delta}-(J-4 U+4\mu)n_D^{\, i,\delta}
% +(4\mu-J)n_V^{\, i,\delta} \right] \, ,\nonumber\\
 & & -\sum_{i,\delta=\delta_x,\delta_y}\left[J n_A^{\, i,\delta}+(\frac{J}{4}- U+\mu)n_D^{\, i,\delta}
 -(\mu-\frac{J}{4})n_V^{\, i,\delta} \right] \, ,\nonumber
\eea
where $\vec{\Delta}=(\Re \Delta,\Im\Delta)$, $\vec{\pi}^{\alpha}=
(\Re \pi^{\alpha},\Im\pi^{\alpha})$ (see appendix~\ref{App1}).
The value of $U$ has to be fine-tuned in order to obtain $SO(5)$ symmetry on a single bond. 
The resulting constraint is $\mu=\frac{J}{4}$ as before, but in
addition $U=\frac{J}{2}$. Note that this
is more restrictive than the local constraint for the ladder model 
\cite{SO5ladder}, which leaves two free parameters. Since we only consider
states of paired electrons, this model has fewer local $SO(5)$ invariants than
the ladder.

To establish $SO(5)$ symmetry of the inter-pair interactions, one now has to take
$t_1=-t_2=\frac{1}{8}$. After subtraction of ${\cal H}_1$, this
yields the Hamiltonian
\bea
{\cal H}_{\rm SO(5)} & = & -\frac{1}{4}\sum_{<l,m>}\vec{N}_{l}\cdot
\vec{N}_{m} -J\sum_{l}n_{A}^l \, ,
\label{Hso5}
\eea
where $l$ and $m$ run over the square lattice spanned by the long bonds
(dotted lines in Fig.\ref{fig het rooster}). The unprojected $SO(5)$ superspin 
$\vec{N}$ is given by Eq.~(\ref{N}).
The local $SO(5)$ invariant $n_A$ is related to the length of the superspin
through $\vec{N}^2=1+4 n_A$. 

Note that Eq.(\ref{Hso5}) 
is not the most general $SO(5)$-symmetric Hamiltonian which
could be formulated. In principle, there can be an additional term of the form
\bea
\lambda\sum_{<l,m>}\sum_{a<b}L_{ab}^lL_{ab}^m
& = &  \lambda \sum_{<l,m>}\left[ \sum_{\alpha}\vec{\pi}^{\alpha}_{l}\cdot\vec{\pi}^{\alpha}_{m}
 \right. \nonumber \\
 & & \left.
+ \vec{S}_{i,\delta_1}\cdot\vec{S}_{i,\delta_2}+Q_{i,\delta_1}Q_{i,\delta_2}
\right] \, ,
\eea
which is also an $SO(5)$ invariant. The
charge-charge interaction $Q_{i,\delta_1}Q_{i,\delta_2}$ was omitted from the
present analysis and a term of this form therefore does not appear at the
symmetric point.

The projected $SO(5)$ (p$SO(5)$) symmetry at $U\rightarrow 
\infty$ evolves from the true $SO(5)$ symmetric point at fine-tuned $U$ in the 
following way. Let us assume we have  $t_1=-t_2=t$ and $U=\frac{J}{2}+\bar{U}$. 
\bea
{\cal H} & = & -\sum_{<l,m>}\left(
2t \vec{\Delta}_{l}\cdot\vec{\Delta}_{m}+
\frac{1}{4}\vec{\St}_{l}\cdot \vec{\St}_{m}\right) 
\nonumber \\
& & 
+\sum_{l}\left(\bar{U}n_{D}^l-J n_{A}^l\right) \, .
\eea
The superspin has no preferred global direction for $\bar{U}=0$, $t=\frac{1}{8}$.
Since the AF groundstate does not have a $|D\ra$ component, while the SC does, a small positive $\bar{U}$ will flop the superspin to the AF direction. 
The energy-difference between the AF and the SC
state can be compensated by an increase in $t$. For $\bar{U}\rightarrow
\infty$, this procedure shifts the superspin-flop point from $t=\frac{1}{8}$ 
to $t=\frac{1}{4}$, with the SC groundstate now having $\la n_D\ra =0$. The shift in
$t$ is accounted for by the different relative normalization of $\vec{\Delta}$ 
and $\vec{\St}$ in the definitions of $\vec{N}_P$ and $\vec{N}$.

\section{Collective modes}

In the above, it was shown that the inter-sublattice hopping $t_2$, which 
couples the spin- and charge-dynamics in our model, plays a crucial role in
establishing the $SO(5)$-symmetry. This symmetry only emerges at the mean-field
level for a specific value of $t_2$. To further investigate the role of this
hopping process, the collective modes in the antiferromagnetic and spin-disordered phases are analyzed. For $t_2=0$, it is found that a decoupled
spin/charge perspective suffices to understand these modes, as one would
expect. In this case, the superconductivity does not
affect the collective spin-modes of the system.

This changes for non-zero $t_2$. Although the dispersion-relations do not 
change qualitatively, the interpretation of the modes does. Most strikingly, the
gapped spin-magnon mode of the Singlet SC phase acquires a $\pi$-mode component.
This mode softens at the transition to the AFSC phase and becomes a pure,
acoustic $\pi$-mode as the system is tuned towards p$SO(5)$-symmetry.

The mode-spectrum for systems with p$SO(5)$-symmetry was analyzed by
Zhang {\em et al.} \cite{pSO5}. We reproduce their results for the present
model and investigate the influence of further $SO(5)$ symmetry-breaking
terms.
%, establishing a difference between
%projected $SO(5)$-symmetry and mean-field $SO(5)$-symmetry (although both
%only become true symmetries in the $d\rightarrow\infty$ limit).

%The mode-spectrum in the antiferromagnetic and spin-disordered phases is
%analyzed. It is discussed to what extend this spectrum can be understood
%rom a completely decoupled spin/charge perspective, which is found to be  
%correct far away from the spin-disordering transition. The interplay between %charge- and
%spin-dynamics, driven by the $t_2$ hopping process, is shown to play an
%important role near this transition in the superconducting phase. 
%The $\pi$-modes of $SO(5)$ theory
%appear in the mode-spectrum as a result of the $t_2$-process. 
%The results obtained in \cite{pSO5} for the mode-spectrum of 
%projected $SO(5)$ models (a quadratic, gapless phase-mode; mixing
%of $\pi$-modes and spin-waves, resulting in a doping-independent Goldstone-mode
%velocity) are reproduced
%for our model.
% for the case that projection onto the lower Hubbard band is
%the only cause of $SO(5)$ symmetry-breaking.
%The influence of
%further symmetry-breaking effects
%, specifically: a more realistic
%choice for the spin-spin interactions and tuning away from $t_2=-\frac{1}{4}$ 
%while remaining at the tricritical point, 
%on the p$SO(5)$ mode-spectrum 
%is investigated.

\subsection{Random phase approximation}

The collective modes of the system are studied in the random phase
approximation (RPA) \cite{RPA}. They are obtained from the equations of
motion of the operators $G_{\alpha\beta}$, which are given by
\beq
\imath \partial_t G_{\alpha\beta}^{i,\delta}= {[}G^{i,\delta}_{\alpha\beta},
{\cal H}{]} \, .
\eeq
The commutator in this expression contains products of operators
on different bonds ($i$,$\delta$). These products are decoupled in 
a mean-field fashion, yielding
a set of coupled linear differential equations.
 After a transformation to frequency-
and momentum-space, it takes the form
\beq
\omega G_{\alpha\beta}(\vec{k},\omega)=\sum_{\alpha'\beta'}M_{\alpha\beta}^{\alpha'\beta'}(\vec{k})G_{\alpha'\beta'}(\vec{k},\omega)\, .
\label{dynMat}
\eeq
The dispersion-relations of the collective modes are obtained from the 
eigenvalues of the dynamical matrix $M$, while its eigenvectors give the
operators which generate these modes. 

% RPA is not a controlled approximation, 
%but it is generally reasonable well inside an ordered phase. It becomes
%particularly poor near phase-transition, which is unfortunate, since this
%is the region of most interest to us here. In the following, we will be
%discussing the RPA-spectrum {\em at} the AFSC to Singlet SC phase-transition.  
%The results there should be viewed more as relating to the mode-spectrum inside
%the Singlet SC phase but close to the phase-transition, where the approximation
%may be valid at long-wavelengths, rather than about the true mode-spectrum
%exactly at the phase-transition.

There is a problem with the above decoupling in the spin-ordered
phases. 
As was discussed in section \ref{ssectie trans fluc}, the low-energy 
fluctuations of the spin-system behave differently at large $J_F$ and near the spin-disordering 
transition. For $J_F\ll 1$, the spins around each square plaquette are 
locked into a symmetric state, forming one spin-2 object, and the low-energy
deformations of the spin-state are localized on the antiferromagnetic 
bonds. The above decoupling,
which cuts across the ferromagnetic bonds, then becomes very poor.
Near the spin-disordering transition, the spins are rigidly coupled
along the antiferromagnetic bonds and the low-energy  transversal 
fluctuations are localized on the ferromagnetic bonds. In this case, the
decoupling works well. 

The cross-over to spin-2 behavior at large $J_F$ is driven by the 
${\cal H}_1$ spin-spin interaction term, Eq.(\ref{H1}).
To avoid it, we calculate the mode-spectrum of the Hamiltonian from which this
term has been subtracted. The resulting dynamical matrices are listed in
appendix~\ref{ssectie dynmat}. 
Subtracting ${\cal H}_1$ makes no difference for the spin-disordered
phases, but does change the results in the AF and AFSC phase (we discuss
in what way).
The ${\cal H}_1$ term breaks $SO(5)$ symmetry,
though retaining it at the mean-field level. As a result, the
model which we study in RPA has a projected $SO(5)$ symmetry 
at the tricritical point with fine-tuned
$t_2$.

%In this case, we indeed reproduce
% the results obtained by Zhang {\em et al.} for 
%p$SO(5)$ models. These results are discussed below. The additional symmetry-breaking ${\cal H}_1$, which we omitted, has no effect on the RPA mode-spectrum in the 
%spin-disordered phases, although we do expect it to affect some
%of the p$SO(5)$ properties in the AFSC phase. This is also discussed.

\begin{figure}[t]
\epsfxsize=.7\hsize
\hspace{.15 \hsize}
\epsffile{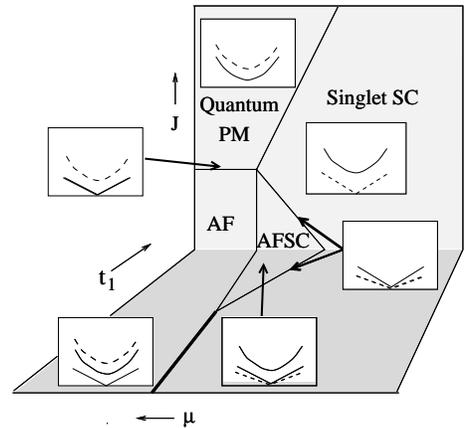}\vspace{8pt}
%\epsffile{proef2.eps}
\caption{Sketch of the mode-spectrum in the antiferromagnetic and
spin-disordered phases.
The dashed lines are pairing modes when gapped (insulating phases) and
phase Goldstone modes when acoustic (superconducting phases).}
\label{fig modesschets}
\end{figure}

\subsection{mode-spectrum}

We briefly discuss the mode-spectrum of the
antiferromagnetic and spin-disordered phases. These results are 
summarized in Fig.~\ref{fig modesschets} .

The quantum paramagnet, for which both spin- and gauge-symmetry are 
unbroken, has no Goldstone modes. Its spectrum consists of a 
gapped three-fold degenerate spin-1 magnon mode and a pairing mode which
is also gapped. The pairing gap closes at the transition to the singlet
superconductor.
Precisely at the transition, this mode has a quadratic dispersion.
As the hole-pair density is increased from zero, it acquires a finite
velocity, becoming the phase Goldstone mode of the superconducting state. 
This is the behavior expected at a dilute
boson transition \cite{boek}, of which this is a particular example
(where the
hole-pairs are the dilute bosons). The three-fold degenerate mode of the quantum paramagnet remains gapped 
through the transition to the 
superconductor.

The insulating antiferromagnet has a two-fold degenerate spin-wave
mode. In addition, it has a gapped mode related to spin-amplitude modulations
and a gapped pairing mode. The spin-amplitude mode becomes degenerate
with the acoustic spin-wave modes at the transition
to the quantum paramagnet, where they turn into the spin-1 magnon triplet.
This transition is in the same class as the spin-disordering transition of
the Heisenberg bilayer model \cite{CM}.
The pairing gap closes at the transition to the antiferromagnetic superconductor, where it becomes the acoustic phase Goldstone mode at 
finite dopings. The spin-amplitude mode remains gapped through the
insulator to superconductor transition, but becomes degenerate with the acoustic spin-wave modes at the subsequent transition to the singlet superconductor.
At this transition line, the system therefore has four acoustic modes, 
of which three are degenerate. 

%\subsection{The $t_2$-process}
\subsection{$\pi$-modes} 

The mode-spectrum as outlined above can be understood entirely from a 
decoupled perspective of the spin- and the phase-sector. At the transition
line from insulator to superconductor, the pairing mode softens
and continuously acquires a finite velocity. At the transition from a
phase with AF order to a spin-disordered phase, the spin-amplitude mode 
becomes degenerate with the acoustic spin-wave modes. These two effects
combined yield the described behavior, and in particular the occurrence of
four acoustic modes at the AFSC to Singlet SC transition. This decoupled
perspective is correct for $t_2=0$. In this case, the gapped modes in the
Singlet SC state are indeed spin-1 magnons, as they are in the quantum paramagnet, while the gapped mode in the AFSC is indeed a pure spin-amplitude mode, as it is for the AF insulator.

The $t_2$-process, however, provides a coupling between the movement of the
hole pairs and the dynamics of the spin-system. This coupling changes the 
nature of the gapped modes in the 
superconducting phases with respect to those in their insulating
parent-phase. In the AFSC phase,
the spin-amplitude mode is mixed with fluctuations between the hole-pair
and the zero-magnetization electron-pair states. In the Singlet SC phase,
$\pi$-modes (hole-pair to triplet fluctuations) are mixed
into the spin-1 magnons. 

Close to the transition from the Singlet to the antiferromagnetic SC, the 
gapped mode becomes degenerate with the acoustic spin-wave modes and
the $t_2$-process begins to affect the low-energy physics of the
system. From the
analogy with the spin-disordering transition in the insulating phase, one
would expect to find a three-fold degenerate acoustic magnon-mode at this
transition (in addition to the phase mode). 
Instead, the RPA-analysis yields an eigenvector
\bea
 & & 4 t_2 \sqrt{J-2}\left(G_{0V}-G_{V0}\right) \nonumber \\
 & & + (J-8 t_2)(1-2 t_1+2 t_2)\left(G_{0A}-G_{A0}\right) \, ,
\label{demode}
\eea
which has both a magnon and a $\pi$-mode component. Note that
this result does not change if the more natural spin-spin interactions,
with ${\cal H}_1$, 
are used, since ${\cal H}_1$ does not affect the collective modes in the
singlet SC phase.

For $J\rightarrow 2$, the spin-disordering line in the 
superconducting phase approaches half-filling. In this case,  
Eq.(\ref{demode}) becomes a pure magnon-mode, which is the result
expected for the insulating phase. The same
eigenvector is found for $t_2=0$, which implies that the 
transition in the superconducting phase is, for that case, 
indeed of the same type as for the insulators. The $\pi$-modes
are mixed in for finite $t_2$. We find a pure $\pi$-mode 
for $t_2=J/8$ and $t_1-t_2=1/2$. The first condition is satisfied at the
point where the AFSC, triplet SC and singlet SC meet, at $n\rightarrow 0$
(see Fig.\ref{f5}). This is related
to the fact that the $n=0$-state and the triplet SC, which
are related by a $\pi$-rotation, become at that point
degenerate in energy. 
The second condition is of more interest: it is fulfilled if the 
singlet-pair hopping-process and the N\'eel-moment interaction 
enter the Hamiltonian in the 
projected $SO(5)$-symmetric form 
$\vec{N}_{\rm p}^{i,\delta_1}\cdot \vec{N}_{\rm P}^{i,\delta_2}$.
This is of course the case at the p$SO(5)$-point, which implies that the
Singlet SC phase at this point has the mode-content expected from $SO(5)$-theory: 
4 acoustic modes, of which one is a phase-mode and three are $\pi$-modes. 
It is shown in Ref.\cite{pSO5} that this is generally the case for 
systems with a projected $SO(5)$ symmetry in the SC phase.
The symmetry breaking due to the projection onto the lower Hubbard band
shows up in the RPA mode-spectrum by a different velocity for the
phase- and the $\pi$-modes:
\bea
v_{\rm phase}^{SC, pSO(5)} & = & \sqrt{\frac{4-J^2}{8}} \, ,
\nonumber \\
v_{\pi}^{SC, pSO(5)} & = & \frac{2+J}{4\sqrt{2}} \, .
\eea 
The two modes become degenerate at $J$=6/5 ($n$=4/5), but this point does not
seem to have any special significance,

The condition $t_1-t_2=1/2$ can also be satisfied at the AFSC to Singlet 
SC transition away 
from the point with mean-field $SO(5)$ symmetry. In this case,
there are
additional terms which break the mean-field $SO(5)$ symmetry, since they
tune the system away from the tricritical point, 
but which do not affect the RPA-modes.
%The modes in the Singlet SC phase exhibit a similar insensitivity 
%to the spin-spin interaction term ${\cal H}_1$, which provides additional
%$SO(5)$-symmetry breaking {\em without} affecting the mean-field groundstate.

\subsection{Projected $SO(5)$ symmetry}

In Ref.\cite{pSO5}, the mode-spectrum at the p$SO(5)$-point
was studied for a general direction of the superspin. Two striking 
results were obtained. In the first place, the system has a two-fold degenerate
acoustic mode, whose velocity is independent of doping  (i.e., independent
of the direction of the $SO(5)$-order 
parameter). Secondly, the phase Goldstone mode is not acoustic, but gapless
with a quadratic dispersion. It was argued that this last effect is caused by the 
infinite compressibility of the system at the p$SO(5)$ point, where
$\partial\la n\ra/\partial \mu$ diverges.

Both results are reproduced in our model. Since the second result is
related to the infinite compressibility rather than to $SO(5)$-symmetry, it
always occurs at the tricritical point, also if we tune away from
$t_1=-t_2=\frac{1}{4}$ while keeping $t_1=t^*, \mu=\mu_c$. By the
same argument, this effect will not disappear if the ${\cal H}_1$-term 
is added to the Hamiltonian, since this does not affect the 
mean-field phase-diagram. 

The first result is very sensitive to
perturbations. As soon as the tricritical point is tuned away from 
$t_1=-t_2=\frac{1}{4}$, the velocity of the acoustic modes becomes 
$n$-dependent. Also the addition of ${\cal H}_1$ to the Hamiltonian
destroys this effect. This
can be seen by calculating the spin-wave velocity from $c=\sqrt{\rho_s/\chi_{\perp}}$, using the results obtained in
section \ref{ssectie trans fluc}, and evaluating it at the mean-field $SO(5)$-point. This yields
\beq
v_s^{MF\,SO(5)}=\sqrt{\frac{(2+J)(J+2-2n))(J+1-n)}{8(J+2 n)}}\, ,
\label{vsMF}
\eeq
which has an $n$-dependence. 
For the model without ${\cal H}_1$, the stiffness is given by
(see Eq.(\ref{rhos pairs},\ref{rhodef}))
\bea
\rho_s^{{\cal H}-{\cal H}_1 } & = & \frac{2}{N\delta \phi^2}\Delta E(\alpha=0)   \\
 & = & n(1-n)(t_1+t_2)(1-\cos 2\chi)+\frac{n^2}{4}(1-\cos^2 2\chi)] \, ,
\nonumber
\eea
where $n$ and $\chi$ have the mean-field values listed in table 1.
The susceptibility is obtained by multiplying the $J_F$-term in the mean-field
energy Eq.(\ref{Emf}) with a factor $\cos^2\tphi^y$, adding the magnetic
field term Eq.(\ref{Emagn}), and minimizing with respect to $\tphi^y$.
We obtain
\beq
\chi_{\perp}^{{\cal H}-{\cal H}_1}=\frac{2n(1-\cos2\chi)}{J-8 t_2(1-n)+n(1-\cos 2\chi)} \, .
\eeq
At the p$SO(5)$ point, this stiffness and susceptibility 
reproduce the RPA result $v_s=(2+J)/(4\sqrt{2})$, which is independent of 
doping. Both the phase-ordering  and the spin-ordering energy contribute to
the spin-wave velocity in the AFSC phase. As one approaches the p$SO(5)$ 
point, the doping-dependence of the contribution of the
spin-ordering energy is precisely compensated by the opposite 
doping-dependence of the phase-ordering contribution. Note that the model 
{\em with} ${\cal H}_1$
yields the same $v_s$ at the transition to the Singlet SC, where $n=(2+J)/4$,
see Eq.~(\ref{vsMF}). This
demonstrates the insensitivity of the RPA mode-spectrum in the Singlet SC
phase to the specific form of the spin-spin interactions.
  
%The doping-independence of the Goldstone-mode velocity at the p$SO(5)$ point
%is not generally present at the mean-field $SO(5)$ point. Therefore, 
%although mean-field and projected $SO(5)$-symmetry both become true symmetries
%only for $d\rightarrow\infty$, they are not the same.
 
At the p$SO(5)$ point of our model, the acoustic modes are no longer pure spin-wave, 
but a combination of spin-wave and $\pi$-mode. Their eigenvector is given by
\beq
\sqrt{1-n}(G_{1V}+G_{V-1})+\sqrt{n-n_c}(G_{10}+G_{0-1}) \, ,
\eeq
which starts out as a spin-wave at half-filling, but crosses over to
a $\pi$-mode at $n=n_c$. This agrees with Zhang {\em et al.}'s interpretation
of the doping-independent velocity in terms of the projected $SO(5)$ symmetry \cite{pSO5}.

\section{Summary}
We have introduced a strong coupling model for spin-ordering and
superconductivity. The microscopic building blocks of this model are
nearest-neighbor electron pairs. The spatial structure of these 
pairs gives rise to d-wave superconductivity. At the same time, it
allows the pairs to have a non-zero uniform or staggered
magnetic moment. In order to
avoid problems related to dimer-type spatial correlations between the
pairs, the model is formulated on a 1/5-depleted lattice. A rich
mean-field phase-diagram is obtained, exhibiting in particular a
phase which is at the same time an antiferromagnet and a superconductor. 
The second order
lines separating this phase from the antiferromagnetic insulator and 
the spin-disordered superconductor end at a tricritical point, where
the antiferromagnet to superconductor phase-transition becomes first
order. By mapping the spin-sector in the antiferromagnetic phases onto
a non-linear sigma model, the main corrections to the mean-field phase-diagram
have been obtained. 

For a specific value of one of the model parameters, a mean-field 
$SO(5)$ symmetry between the antiferromagnetic and superconducting
order-parameter appears to be realized at the tricritical point. It turns out 
that the model still contains spatial gradient terms which 
break $SO(5)$ symmetry. These can be removed by modifying the spin-spin interactions.
The remaining $SO(5)$ symmetry-breaking is then 
a pure quantum-effect, being related to the operator-algebra rather than
the Hamiltonian.
It is shown that
true $SO(5)$ symmetry can be realized for this model by
allowing double site-occupancy and fine-tuning the Hubbard $U$. The
approximate symmetry at large $U$ is therefore a projected $SO(5)$
symmetry of the kind discussed in Ref.\cite{pSO5}.

We investigated the mode-spectrum using the random phase approximation.
It is found that the inter-sublattice hopping process gives rise to the
appearance of a $\pi$-component in the gapped modes of the Singlet SC
phase. Approaching the point with projected $SO(5)$ symmetry from the
Singlet SC phase, a three-fold degenerate
acoustic $\pi$-mode is found as well as an acoustic phase-mode. The RPA
mode-spectrum then has the properties expected for an $SO(5)$-symmetric
system in the pure superconducting phase, apart from the fact that 
the $\pi$-modes and the phase-mode have different velocities.   

As reported in Ref.\cite{pSO5}, the system at the p$SO(5)$ point
has a gapless phase-mode with a quadratic dispersion, as well as a 
two-fold degenerate acoustic mode whose velocity is independent of 
doping. This acoustic mode crosses over from a pure spin-wave at half-filling 
to a $\pi$-mode at the transition to the Singlet SC phase.
We find that the quadratic phase-mode is a property of the tricritical 
point rather than of the projected $SO(5)$ symmetry. The doping-independent
velocity, however, is a strong signature of projected $SO(5)$ symmetry, which
can be destroyed even by additional symmetry-breaking terms that leave the
mean-field $SO(5)$ symmetry intact. 
%Projected $SO(5)$ symmetry is therefore
%more sensitive to perturbations than mean-field $SO(5)$ symmetry.

{\em Acknowledgements.}  Financial support was provided by the Foundation
of Fundamental Research on Matter (FOM), which is sponsored by the
Netherlands Organization of Pure research (NWO). JZ acknowledges support
by the Dutch Academy of Sciences (KNAW). 

\appendix
\section{The $SO(5)$ algebra.}
\label{App1}
A short overview is given of the   
representation of the $SO(5)$-algebra for this model. 

In the unprojected Hilbertspace, a representation of the $SO(5)$ algebra can be
defined which transforms the superspin $\vec{N}$ as a vector. The superspin is
given by
\bea
\vec{N} & = & \left(\Re\Delta,\vec{\St},\Im\Delta\right)\, ,
\label{N}
\eea
where
\beq
\Delta^{\dagger} = \sqrt{2}\left(G_{DA}-G_{AV}\right) \,
\eeq
and $\Re\Delta=\frac{1}{2}\left(\Delta^{\dagger}+\Delta\right)$, $\Im\Delta
= \frac{1}{2\imath}\left(\Delta^{\dagger}-\Delta\right)$. The generators
of the $SO(5)$-algebra satisfy the commutation relation
\beq
{[}L_{ab},L_{cd}{]} = \imath \left(\delta_{ac}L_{bd}+\delta_{bd}L_{ac}-\delta_{ad}L_{bc}-\delta_{bc}L_{ad}\right)
\, ,\label{so5-algebra}
\eeq
where the indices take the values 1 through 5. The $L_{ab}$ are anti-symmetric
under an interchange of $a$ and $b$. They are given by \cite{Zhangscience}
\beq
L_{ab} = \left(\begin{array}{ccccc} 0 & & & & \\
				2\Re\pi_x & 0 & & & \\
				2\Re\pi_y & -S^z & 0 & & \\
				2\Re\pi_z & S^y & -S^x & 0 & \\
Q & 2\Im\pi_x & 2\Im \pi_y & 
 2\Im \pi_z & 0 \end{array}\right)
\label{Lab}
\eeq
where the $\pi$-operators read $\pi^{\dagger}_{\alpha}=-\frac{1}{2}c_1^{\dagger}\sigma_{\alpha}\sigma_y
c_2^{\dagger}$, with $\vec{\sigma}$ the Pauli matrices \cite{SO5ladder}. Projecting onto the
paired-electron states, we obtain
\bea
\pi_x^{\dagger} & = & \frac{1}{2\imath}\left(G_{D1}-G_{D\,-1}+G_{1V}-G_{-1V}\right)
\, ,\nonumber \\
\pi_y^{\dagger} & = & \frac{1}{2}\left(G_{D1}+G_{D\,-1}-G_{1V}-G_{-1V}\right)
\ ,\nonumber \\
\pi_z^{\dagger} & = & \frac{\imath}{\sqrt{2}}\left(G_{D0}+G_{0V}\right) \, .
\label{pi}
\eea
The charge-operator is given by
\beq
Q=n_D-n_V \, .
\label{Q}
\eeq

It can be checked that $\vec{N}$ indeed
transforms as a vector under this $SO(5)$-algebra:
\beq
{[}L_{ab},N_c{]} = \imath\left(\delta_{ac}N_b-\delta_{bc}N_a\right)\, ,
\label{vector}
\eeq
and furthermore that 
\beq
{[}N_a,N_b{]} = \imath L_{ab} \, .
\label{NN}
\eeq

\section{Dynamical matrices}
\label{ssectie dynmat}

%After subtraction of ${\cal H_1}$, the operators in the 
%Hamiltonian Eq.~(\ref{Ham}) can be
%labelled by their position on the square lattice spanned by the long bonds,
%fig.~\ref{fig het rooster}. We had already absorbed a staggering factor
%into the singlet-state, which means that the antiferromagnetic spin-state
%is uniform in terms of the operators $\vec{\St}$, Eq.(\ref{St}). Absorbing
%a d-wave staggering-factor into the hole-pair state $|V_{i,\delta}\ra $, 
%the d-wave SC state is uniform in terms of the operators $G_{\alpha V}$. 
%Working in terms of the staggered operators, it is not necessary to introduce
%different spin- and phase-sublattices in our analysis.

A staggering factor for the antiferromagnetic spin- and the d-wave phase-order
has been absorbed into the operators $G_{\alpha\beta}$. After subtraction 
of ${\cal H}_1$, the Hamiltonian takes the form of a model on the square
lattice, where the operators $G_{\alpha\beta}$ act on the states on the 
lattice sites. The Singlet dSC, AF dSC, quantum paramagnet and AF insulator mean-field states are all uniform in terms of these operators. It is therefore 
not necessary to introduce a multi-sublattice structure. The different 
modes in terms of the real (non-staggered) operators are simply related
to the ones obtained here by a shift in $k$-space.

The operators $G_{\alpha,\beta}$ separate into three sets. Each operator
couples only to operators in the same set through its
RPA equation of motion. One set is formed by the raising operators
$\{G_{1V},G_{1A},G_{10},G_{V-1},G_{A-1},G_{0-1}\}$, another by the
lowering operators, which are related to this set by hermitian conjugation.
The third set contains the operators which act only on the 
zero-magnetization states: $\{G_{AV},G_{0V},G_{A0},G_{VA},G_{V0},G_{0A},n_A,n_0,n_V\}$.
%The operators which are in none of these sets play no role in the RPA collective modes
%of the system.

The dynamical matrix of the raising operators has the form
\beq
M_R=\left(\begin{array}{cc} A^T & B^T \\ -B^T & -A^T \end{array}\right) \, ,
\eeq
where $A$ and $B$ are the $3\times3$ matrices
\beq
A= \left(\begin{array}{ccc} 4\Sigma_t s_{\theta}^2 \gamma_k+\mu-\frac{J}{4} & -2s_{2\theta}c_{\chi}\left(\Delta_t-\frac{\gamma_k}{4}\right) & \Sigma_t 
2s_{2\theta}  s_\chi \\
2 s_{2\theta}c_{\chi}(\Sigma_t \gamma_k-\Delta_t) & c_{\theta}^2c_{\chi}^2\gamma_k-J 
& c_{\theta}^2s_{2\chi} \\
2 \Sigma_t s_{2\theta}s_{\chi}(1-\gamma_k) & \frac{1}{2} c_{\theta}^2s_{2\chi}(2-\gamma_k)
& 0 
\end{array}\right) \, ,
\eeq

\beq
B=\left(\begin{array}{ccc} 0 & \frac{1}{2}s_{2\theta}c_{\chi}\gamma_k & 0 \\
0 & c_{\theta}^2c_{\chi}^2 \gamma_k & 0 \\
0 & -\frac{1}{2}c_{\theta}^2s_{2\chi}\gamma_k & 0 
\end{array}\right) \, ,
\eeq
where we have used the notation
\bea
s_x & = & \sin x \; ; \; c_x=\cos x \, , \nonumber \\
\Sigma_t & = & t_1+t_2 \; ; \; \Delta_t=t_1-t_2 \, ,\\
\gamma_k & = & \frac{1}{2}(\cos (\vec{k}\cdot\vec{e}_1) + \cos (\vec{k}\cdot \vec{e}_2) ) \, ,
\nonumber
\eea
with $\vec{e}_1$ and $\vec{e}_2$ the basis-vectors of the square lattice spanned by the
long bonds, Fig.~\ref{fig het rooster}. The angles $\chi$ and $\theta$ are the ones 
appearing in the mean field energy Eq.(\ref{Emf}). In the insulating phases, $\theta$ vanishes, while $\chi$ is equal to zero 
in the spin-disordered phases.

The dynamical matrix of the lowering modes is the same as $M_R$, apart from a minus sign.
The last set has a dynamical matrix
\beq
M_0=\left(\begin{array}{ccc} C & D^T & E^T \\ -D^T & -C & -E^T \\ F & -F & 0 \end{array}\right) \, ,
\eeq
which consists of the $3\times 3$ matrices
\bea
C_{11} & = & -4 \Delta_t(c_{\chi}^2 c_{\theta}^2-s_{\theta}^2)\gamma_k+\mu+\frac{3}{4}J \nonumber \\
C_{12} & = & 2 \Delta_t c_{\theta}^2s_{2\chi}\gamma_k-c_{\theta}^2s_{2\chi} \nonumber \\
C_{13} & = & -2 s_{2\theta}s_{\chi}(\Delta_t\gamma_k-\Sigma_t) \nonumber \\
C_{21} & = & 2 \Sigma_t c_{\theta}^2s_{2\chi} \gamma_k-c_{\theta}^2s_{2\chi}  \nonumber \\
C_{22} & = & -4 \Sigma_t(c_{\theta}^2s_{\chi}^2-s_{\theta}^2)\gamma_k+\mu-\frac{J}{4} \nonumber \\
C_{23} & = & 0 \nonumber \\
C_{31} & = &  2 \Sigma_t s_{2\theta} s_{\chi}-\frac{1}{2}s_{2\theta}s_{\chi} 
\nonumber \\
C_{32} & = & \frac{1}{2} s_{2\theta} c_{\chi} \gamma_k \nonumber \\
C_{33} & = & -c_{\theta}^2c_{2\chi}\gamma_k+ J
\eea
\beq
D=\left(\begin{array}{ccc} 0 & 0 & \frac{1}{2}s_{2\theta}s_{\chi}\gamma_k \\
 0 & 0 & 2 \Delta_t s_{2\theta}c_{\chi}
-\frac{1}{2}s_{2\theta}c_{\chi}\gamma_k \\
0  & -2s_{2\theta}c_{\chi}(\Delta_t-\Sigma_t\gamma_k)   & c_{\theta}^2c_{2\chi}\gamma_k \end{array}\right)
\eeq
\beq
E=\left(\begin{array}{ccc} 
-2\Delta_t s_{2\theta}c_{\chi}(1-\gamma_k) & 0 & c_{\theta}^2s_{2\chi} \\
0 & 2 \Sigma_t s_{2\theta}s_{\chi}(1-\gamma_k) & -c_{\theta}^2s_{2\chi} \\ 
2\Delta_t s_{2\theta}c_{\chi}(1-\gamma_k)
 & -2 \Sigma_t s_{2\theta}s_{\chi}(1-\gamma_k)  & 0 \end{array}\right)
\eeq

\beq
F=\left(\begin{array}{ccc}-2 \Delta_t s_{2\theta}c_{\chi} & 0 & c_{\theta}^2s_{2\chi} \\
 0 & 2 \Sigma_t s_{2\theta}s_{\chi} & -c_{\theta}^2s_{2\chi} \\
2 \Delta_t s_{2\theta}c_{\chi} & -2 \Sigma_t s_{2\theta}s_{\chi} & 0 \end{array}\right)
\eeq

For the Singlet SC phase ($\chi=0$), the operators $G_{\alpha\beta}$ with $\alpha$ and $\beta$ referring
to triplet states decouple from the equations of motion, since there is no longer 
a triplet component in the mean-field groundstate. This leaves the raising set
$\{G_{1V},G_{1A},G_{V-1},G_{A-1}\}$ and the hermitian conjugate lowering set. The set of 
zero-magnetization operators splits into the transversal set
$\{G_{0V},G_{0A},G_{V0},G_{A0}\}$ and the longitudinal set $\{G_{AV},G_{VA},n_A-n_V\}$. 
The dynamical matrices of the first three sets contain the three-fold degenerate $\pi$- and 
spin-1 magnon modes. The fourth set contains the phase and the pairing mode.

In the quantum paramagnet phase ($\chi=\theta=0$), the first three sets further simplify. Since
the $\pi$-operators do not refer to the spin-singlet groundstate, only the
spin-1/charge-0 operators are left in these sets. The phase-mode $n_A-n_V$
disappears from the last set.

\end{multicols}
\end{document}